\documentclass[twocolumn,english,aps,pra,showpacs,floats,footnoterule,superscriptaddress]{revtex4}
\usepackage[T1]{fontenc}
\usepackage[latin9]{inputenc}
\usepackage{amsmath}
\usepackage{graphicx}
\usepackage{amssymb}

\usepackage{amssymb,amsmath}
\usepackage[usenames]{color}
\usepackage{graphicx}

\DeclareMathOperator{\cosech}{cosech}

\begin{document}

\title{Non-local pair correlations in the 1D Bose gas at finite temperature}

\author{P.~Deuar}

\email{piotr.deuar@lptms.u-psud.fr}
\affiliation{Laboratoire Physique Th\'{e}orique et Mod\`{e}les Statistique, Universit\'{e}
Paris-Sud, CNRS, 91405 Orsay, France}

\author{A. G. Sykes}

\affiliation{ARC Centre of Excellence for Quantum-Atom Optics, School of Physical
Sciences, University of Queensland, Brisbane, QLD 4072, Australia}

\author{D. M. Gangardt}

\affiliation{School of Physics and Astronomy, University of Birmingham, Edgbaston,
Birmingham B15 2TT, United Kingdom}

\author{M. J. Davis}

\affiliation{ARC Centre of Excellence for Quantum-Atom Optics, School of Physical
Sciences, University of Queensland, Brisbane, QLD 4072, Australia}

\author{P. D. Drummond}

\affiliation{ARC Centre of Excellence for Quantum-Atom Optics, Centre for Atom
Optics and Ultra-fast Spectroscopy, Swinburne University of Technology,
Melbourne, VIC 3122, Australia}

\author{K. V. Kheruntsyan}

\affiliation{ARC Centre of Excellence for Quantum-Atom Optics, School of Physical
Sciences, University of Queensland, Brisbane, QLD 4072, Australia}

\date{\today }
\begin{abstract}
The behavior of the spatial two-particle correlation function is surveyed
in detail for a uniform 1D Bose gas with repulsive contact interactions
at finite temperatures. Both long-, medium-, and short-range effects
are investigated. The results span the entire range of physical regimes,
from ideal gas, to strongly interacting, and from zero temperature
to high temperature. 
We present perturbative analytic methods, available at strong and
weak coupling, and first-principle numerical results using imaginary
time simulations with the gauge-$P$ representation in regimes where
perturbative methods are invalid. Nontrivial effects are observed
from the interplay of thermally induced {bunching} behavior versus
interaction induced {antibunching}.
\end{abstract}


\maketitle

\section{Introduction}

The study of two-body correlations has a long history dating back
to the $1956$ experiment of Hanbury Brown and Twiss (HBT)~\cite{hbt-expt}.
The HBT experiment set out to measure the intensity of light coming
from a distant star, at two nearby points in space. The fluctuations
in the intensities were shown to be strongly correlated in spite of
the thermal nature of the source. In more recent times, experimental
progress in the field of ultra-cold atomic gases has provided the
opportunity to examine similar correlations in systems of cold atoms
(as opposed to photonic systems). The large thermal de Broglie wavelength
in a cold gas means the correlations occur on length scales large
enough to be resolved using current detectors. A pioneering experiment
of this kind involving a cloud of cold Neon atoms, was carried out
by Yasuda and Shimizu~\cite{yasuda-shimizu} as early as $1996$.
A more comprehensive study was undertaken during $2005-2007$ in Refs.~\cite{schellekens,jeltes},
where the two particle \emph{bunching} phenomena associated with Bose
enhancement (when metastable $^{4}$He$^{\ast}$ atoms were used)
was juxtaposed with the \emph{antibunching} behavior present in a
system of fermions (when $^{3}$He$^{\ast}$ atoms were used). In
all of the above cases the measured correlations were completely described
by the statistical exchange interaction between particles in an \emph{ideal}
gas.

The behavior of strongly \emph{interacting} systems poses some of
the most difficult questions confronting current theoretical studies
in many-body physics. In this paper we discuss how our simple understanding
of two-body correlations in an ideal gas can be radically altered
in the presence of interactions. To demonstrate this we calculate
the normalized pair correlation function \begin{equation}
g^{(2)}(r)=\langle\hat{\Psi}^{\dagger}(0)\hat{\Psi}^{\dagger}(r)\hat{\Psi}(r)\hat{\Psi}(0)\rangle/n^{2}\label{eq:g2}\end{equation}
 in a homogeneous repulsive one-dimensional (1D) Bose gas \cite{liebliniger,lieb2}
at finite temperature over a wide range of interaction strengths.
{In Eq. \eqref{eq:g2}, $\hat{\Psi}(x)$ is the field operator, and
$n=\langle\hat{\Psi}^{\dagger}(x)\hat{\Psi}(x)\rangle$ is the linear
1D density}. Physically, $g^{(2)}(r)$ quantifies the conditional
probability of detecting a particle at position $r$, given that a
particle has been detected at the origin. Theoretically the 1D Bose
gas model with $\delta$-function interaction is one of the simplest
paradigms we have of a strongly interacting quantum fluid, owing to
its exact integrability
\cite{liebliniger,lieb2,yangyang1,korepin-book,giamarchi-book,gogolin-book}.
In the limit of an infinitely strong interaction it corresponds to a
gas of impenetrable (hard-core) Bosons treated first in Ref.
\cite{girardeau}. It also holds relevance as an experimentally
accessible
system~\cite{bec_low_dim,bec_low_dim2,Greiner-1D-exp,greiner_low_dim_bec,Richard-1D-exp,esslinger_expt1,bill,bloch_expt,weiss_expt,Weiss-g2-1D,esslinger_expt2,Raizen-BEC-box,Bouchoule-density-density,Schmiedmayer-1D-exp,van_Druten}.
Opposite from 2D and 3D, the strongly interacting limit of a 1D system
is achieved in the low density regime. In this regime the wave function of the particles is strongly
correlated and prevents them from being close to each other,
which results in dramatic suppression of 3-body losses. This
allows for the stable creation of strongly interacting 1D Bose
gases.

There has been a substantial amount of previous theory on
correlations of the 1D Bose gas model. The Luttinger liquid approach
provides a method of calculating the long-range asymptotic behavior
in the decay of non-local correlations
~\cite{giamarchi-book,gogolin-book}. Local second- and third-order
correlations in the homogeneous system have been calculated in
Refs.~\cite{Castin,gangardt_T_0,gangardt-correlations,karenprl,Cazalila};
extensions to inhomogeneous systems using the local density
approximation (LDA) are given in Ref.~\cite{karen-pra}. Numerical
calculations at specific values of interaction strength have been
carried out at $T=0$ \cite{untrapped_via_montecarlo} and at finite
temperature \cite{drummond-canonical-gauge}. Similar \emph{nonlocal}
quantities have been calculated for the $T=0$ ground state
\cite{lenard,schultz,untrapped_via_montecarlo,caux_correlations,caux_correlations2,cherny_brand1},
and for finite temperature both numerically
\cite{drummond-canonical-gauge}
and in the strong interaction limit \cite{cherny_brand2}. 
Refs.~\cite{korepin-book,giamarchi-book,gogolin-book,lieb_book,shlyapnikovlecturenotes,Castin04}
contain recent reviews of the physics of the 1D Bose gas problem.

The focus of the present paper is the nonlocal correlation function
{at arbitrary interparticle separations $r$}; we give the details of
analytic derivations of the results discussed in a recent
Letter~\cite{sykes_raizen} and complement them with exact numerical
calculations using the stochastic gauge-$P$ method of
Ref.~\cite{drummond-canonical-gauge,deuar-drummond-2002,drummond-deuar-2003,deuar-drummond-2006,deuar-thesis}.
Experimental proposals to measure nonlocal spatial correlations
between the atoms in a 1D Bose gas have been discussed in
Ref.~\cite{sykes_raizen,accordion}.

The structure of this paper is as follows. In section \ref{sect:liebformalism}
we give a brief review of the physics of a 1D Bose gas, emphasizing
the important parameters which determine the phase diagram. In section
\ref{sect:numerical} we outline the details involved in the application
of the (imaginary time) gauge-$P$ phase space method to the 1D Bose
gas. The more technical details are placed in appendix~\ref{append:numerix}.
This method is capable of obtaining numerical results in the cross-over
regions of the phase diagram, where analytic results are not available.
In sections \ref{sect:nearlyideal}, \ref{sect:weak} and \ref{sect:strong}
we present the results of calculating $g^{(2)}(r)$ in the nearly
ideal gas limit, the weakly interacting limit, and the strongly interacting
limit respectively. The results are obtained from numerical calculations
and analytic perturbation expansions. We describe the details of our
perturbation expansion in each respective section. In section \ref{sect:HTFxover}
we analyze, in detail, the nature of the crossover into the fermionized
Tonks gas regime. Section \ref{sect:numerical-limitations} discusses
the limitations of the numerical method. In section \ref{sect:conclusions}
we give an overview and draw conclusions.

\section{The Interacting Bose gas in 1D}

\label{sect:liebformalism}

We are considering a homogeneous system of $N$ identical bosons in
a 1D box of length $L$ with periodic boundary conditions \cite{liebliniger,lieb2}.
We include two-body interactions in the form of a repulsive delta-function
potential. The second-quantized Hamiltonian of the system is given
by \begin{equation}
\hat{H}=\frac{\hbar^{2}}{2m}\int dx\,\partial_{x}\hat{\Psi}^{\dagger}\partial_{x}\hat{\Psi}+\frac{g}{2}\int dx\,\hat{\Psi}^{\dagger}\hat{\Psi}^{\dagger}\hat{\Psi}\hat{\Psi},\label{Hfull}\end{equation}
where $m$ is the mass and $g>0$ is the coupling constant that can
be expressed via the 3D $s$-wave scattering length $a$ as $g\simeq2\hbar^{2}a/(ml_{\perp}^{2})=2\hbar\omega_{\perp}a$
\cite{olshanii-1d-scattering}. Here, we have assumed that the atoms
are transversely confined by a tight harmonic trap with frequency
$\omega_{\perp}$ and that $a$ is much smaller than the transverse
harmonic oscillator length $l_{\perp}=\sqrt{\hbar/m\omega_{\perp}}$.
The 1D regime is realized when the transverse excitation energy $\hbar\omega_{\perp}$
is much larger than both the thermal energy $T$ (with $k_{B}=1$)
and the chemical potential $\mu$ \cite{karen-pra,interaction-crossover}.
A uniform system in the thermodynamic limit ($N,L\longrightarrow\infty$,
while the 1D density $n=N/L$ remains constant) is completely characterized
\cite{liebliniger,yangyang1} by two parameters: the dimensionless
interaction strength\begin{equation}
\gamma=\frac{mg}{\hbar^{2}n}\end{equation}
and the reduced temperature\begin{equation}
\tau=T/T_{d},\end{equation}
where $T_{d}=\hbar^{2}n^{2}/(2m)$ is the temperature of quantum degeneracy
in units of energy \cite{karenprl}. 

The interplay between these two parameters dictates the dominating
behavior in six physically different regimes. Briefly, these regimes
are:
\begin{itemize}
\item \emph{Nearly ideal gas }regime, where the temperature always dominates
over the interaction strength. This regime splits into two subregimes
defined by $\tau\ll1$ or $\tau\gg1$. In both cases one must have
$\gamma\ll\min\left\{ \tau^{2},\sqrt{\tau}\right\} $.
\item \emph{Weakly interacting }regime, where both the interaction strength
and the temperature are small, but $\tau^{2}\ll\gamma\ll1$. This
regime realizes the well known quasi-condensate phase. Fluctuations
occur due to either vacuum or thermal fluctuations, which defines
two further subregimes, with $\tau\ll\gamma$ or $\tau\gg\gamma$,
respectively.
\item \emph{Strongly interacting }regime, where the interaction strength
is large and dominates over temperature induced effects. This can
occur at high and low temperatures, again defining two subregimes
with $\tau\ll1$ or $\tau\gg1$.
\end{itemize}
The basic understanding of the competition between interaction induced
effects and thermally induced effects was outlined in Ref.~\cite{sykes_raizen}.

Although the model is integrable via the Bethe ansatz, the
cumbersome nature of the eigenstates \cite{Sykes-Davis-Drummond}
inhibits the direct calculation of the nonlocal two-body correlation
function. We therefore use numerical integration in a phase-space
representation, together with perturbation theory in each of the six
regimes. The standard Bogoliubov procedure, applied to
Eq.~\eqref{Hfull} is appropriate in the case of the weakly
interacting regime (see section \ref{sect:weak}). Perturbation
theory in the strongly interacting and nearly ideal gas regimes is
done using the path integral formalism (see sections
\ref{sect:perturbation} and \ref{sect:strong} respectively).


\section{Numerical Stochastic Gauge Calculations}

\label{sect:numerical}

\subsection{Gauge-$P$ distribution}

To evaluate correlations away from the regimes of applicability of
the analytic approximations, we use the gauge-$P$ phase-space method
to generate a stochastic evolution from the simple $T\rightarrow\infty$
limit (where interactions are negligible) down to lower temperatures.
This method gives results that correspond exactly to the full quantum
mechanics using the Hamiltonian (\ref{Hfull}) as the number of averaged
realizations ($\mathcal{S}$) goes to infinity. The gauge-$P$ method
has been described in \cite{deuar-drummond-2002,drummond-deuar-2003,deuar-drummond-2006},
and is covered in greatest detail in \cite{deuar-thesis}, while an
initial application to the 1D Bose gas was presented in \cite{drummond-canonical-gauge}.
Below we give a summary of the derivation for this system, and present
the basic calculation procedure. Some of the more technical details
are given in Appendix~\ref{append:numerix}.

We consider a grand canonical ensemble with mean density $n$, Hamiltonian
(\ref{Hfull}) and inverse temperature given by $\beta=1/k_{B}T$.
When the Hamiltonian commutes with the number operator $\widehat{N}=\int dx\widehat{\Psi}^{\dagger}(x)\widehat{\Psi}(x)$,
as is the case here, the unnormalized density matrix at temperature
$T$ is given by \begin{equation}
\widehat{\rho}_{u}=e^{[\mu(\beta)\widehat{N}-\widehat{H}]\beta},\label{rhou}\end{equation}
where $\mu(\beta)$ is the chemical potential. In this formulation,
$\mu$ can in principle be chosen at will as any desired function
of temperature, thus indirectly determining the density $n(T)$. In
the Schrödinger picture the density matrix is equivalently defined
by an {}``imaginary time\textquotedblright\ master-like equation
\begin{eqnarray}
\frac{\partial\widehat{\rho}_{u}(\beta)}{\partial\beta} & = & \left[\mu_{e}(\beta)\widehat{N}-\widehat{H}\right]\,\widehat{\rho}_{u}(\beta)\notag\\
 & = & \frac{1}{2}\left[\mu_{e}(\beta)\widehat{N}-\widehat{H}\ ,\ \widehat{\rho}_{u}(\beta)\right]_{+}\label{master}\end{eqnarray}
and a simple initial (i.e. $T\rightarrow\infty$) condition \begin{equation}
\widehat{\rho}_{u}(0)=e^{-\lambda\widehat{N}},\label{rho_ic}\end{equation}
with $\lambda=-\lim_{\beta\rightarrow0}\left[\beta\mu(\beta)\right]$
and $\beta$ playing a similar role to time in the Schrodinger equation
for time evolution, apart from a factor of $i$ (hence the name).
The second line of (\ref{master}) follows from the restricted set
of density matrices described by the grand canonical ensemble (\ref{rhou}),
where $\log\widehat{\rho}_{u}$ commutes with $\widehat{\rho}_{u}$.
Note that $\mu_{e}(\beta)$ is a temperature-dependent {}``effective\textquotedblright\
chemical potential \begin{equation}
\mu_{e}=\frac{\partial\lbrack\beta\mu(\beta)]}{\partial\beta},\end{equation}
that is not necessarily equal to $\mu$. The initial condition (\ref{rho_ic})
can then be evolved according to Eq. (\ref{master}) to obtain the
equilibrium state at lower temperatures $\beta>0$. However, in the
density matrix form, this naturally becomes intractable for more than
a few particles.

Phase-space methods such as the gauge-$P$ distribution used here
reduce the computational resources needed to a manageable number.
This is done by deriving a Fokker-Planck equation for a distribution
of phase-space variables that is equivalent to the full quantum
mechanics (\ref{master}), and then in a second step, sampling this
distribution stochastically and evolving the \textsl{samples} with a
diffusive random walk that is equivalent to the Fokker-Planck
equation. The general approach is described in
\cite{positiveP,Gardiner}. The price that is paid for tractable
calculations is a loss of precision that comes about due to the
finite sample size $\mathcal{S}$. Fortunately this uncertainty can
be readily estimated using the Central Limit theorem and scales as
$\sqrt{\mathcal{S}}$.

We utilize the normalized off-diagonal coherent state expansion of
the positive-$P$ distribution \cite{positiveP} because the number
of variables required to describe a sample is linear in the number
of spatial points (tractability) and because it describes all quantum
states with a non-negative real distribution. However, for this investigation
two additional elements are needed. Firstly, the evolution (\ref{master})
does not preserve the trace, so an additional weight variable in the
expansion is needed to keep track of this. Secondly, the evolution
equations for the samples given by a bare weighted positive-$P$ treatment
are unstable and can lead to systematically bad sampling \cite{Gilchrist}.
The complex part of the weight variable allows us to remove these
instabilities using a stochastic gauge as described in \cite{deuar-drummond-2002,drummond-canonical-gauge}.

In practice, the first step is to discretize space into $M$ equally
spaced points in a box of length $L$ with periodic boundary conditions,
on which the fields are defined. There is a lattice spacing of $\Delta x=L/M$
per point. One must make sure that the lattice is fine enough and
long enough to encompass all relevant detail. In practice we check
this by increasing $L$ and, separately, $M$ until no further change
in the results is seen. Having this equivalent lattice, one can expand
the density matrix $\widehat{\rho}_{u}$ as \begin{equation}
\widehat{\rho}_{u}=\int G(\vec{v})\widehat{\Lambda}(\vec{v})\ d^{4M+2}\vec{v},\end{equation}
with a positive \cite{deuar-drummond-2002} distribution $G(\vec{v})$
of the set of $2M+1$ complex phase-space variables, \begin{equation}
\vec{v}=\left\{ \alpha_{1},\dots,\alpha_{M},\alpha_{1}^{+},\dots,\alpha_{M}^{+},\Omega\right\} ,\end{equation}
that describe an operator basis \begin{equation}
\widehat{\Lambda}(\vec{v})=\Omega\otimes_{j=1}^{M}||\alpha_{j}\rangle\langle(\alpha_{j}^{+})^{\ast}||\ e^{-\sum_{j=1}^{M}\alpha_{j}^{+}\alpha_{j}}\label{lambda}\end{equation}
composed of unnormalized (Bargmann) coherent states $||\alpha_{j}\rangle=\exp\left[\alpha_{j}\sqrt{\Delta x}\,\widehat{\Psi}^{\dagger}(x_{j})\right]|0\rangle$
at the $j$-th point at location $x_{j}=(j-1)\Delta x$ and a global
weight $\Omega$. 

The initial condition (\ref{rho_ic}) corresponds to the distribution
\begin{equation}
G_{0}(\vec{v})=\delta^{2}(\Omega-1)\prod_{j=1}^{M}\delta^{2}\left(\alpha_{j}-(\alpha_{j}^{+})^{\ast}\right)\frac{\exp(-|\alpha_{j}|^{2}/\overline{n}_{x})}{\pi\overline{n}_{x}},\label{G0}\end{equation}
where $\overline{n}_{x}=1/(e^{\lambda}-1)=N/M$ is the mean number
of atoms ($N=\langle\hat{N}\rangle$) per spatial point in the initial
$\beta=0$ state. We see that, at least initially, $\alpha^{+}=(\alpha)^{\ast}$
are complex conjugates.

\subsection{Fokker-Planck Equation}

To generate the Fokker-Planck equation (FPE) for $G(\vec{v})$ corresponding
to the master equation (\ref{master}) we use the following differential
identities for the basis operators \begin{subequations} \label{identities}
\begin{eqnarray}
\sqrt{\Delta x}\,\widehat{\Psi}(x_{j})\widehat{\Lambda} & = & \alpha_{j}\,\widehat{\Lambda},\\
\sqrt{\Delta x}\,\widehat{\Psi}^{\dagger}(x_{j})\widehat{\Lambda} & = & \left(\alpha_{j}^{+}+\frac{\partial}{\partial\alpha_{j}}\right)\widehat{\Lambda},\\
\sqrt{\Delta x}\,\widehat{\Lambda}\widehat{\Psi}(x_{j}) & = & \alpha_{j}^{+}\,\widehat{\Lambda},\\
\sqrt{\Delta x}\,\widehat{\Lambda}\widehat{\Psi}^{\dagger}(x_{j}) & = & \left(\alpha_{j}+\frac{\partial}{\partial\alpha_{j}^{+}}\right)\widehat{\Lambda}.\end{eqnarray}
These convert quantities involving the operators $\widehat{\Psi}$,
$\widehat{\Psi}^{\dagger}$ and $\widehat{\rho}_{u}$ to ones involving
only $\widehat{\Lambda}$ and their derivatives.

In what follows it will be convenient to label the $\alpha$ and $\alpha^{+}$
variables as \end{subequations} \[
\alpha_{j}^{(\nu)}=\left\{ \begin{array}{cl}
\alpha_{j}, & \text{ if }\nu=1,\\
\alpha_{j}^{+}, & \text{ if }\nu=2.\end{array}\right.\]
Using (\ref{identities}) on (\ref{master}) one obtains \begin{eqnarray}
\lefteqn{\int\frac{\partial G(\vec{v})}{\partial\beta}\widehat{\Lambda}\, d^{4M+2}\vec{v}=-\int G(\vec{v})}\label{differential}\\
 &  & \times\left\{ \frac{g}{4\Delta x}\sum_{j,\nu}(\alpha_{j}^{(\nu)})^{2}\frac{\partial^{2}}{\partial(\alpha_{j}^{(\nu)})^{2}}+K(\vec{v})\right.\notag\\
 &  & \hspace*{-2em}\left.+\frac{1}{2}\sum_{j}\left[\left(\frac{\partial K(\vec{v})}{\partial\alpha_{j}^{+}}\right)\frac{\partial}{\partial\alpha_{j}}+\left(\frac{\partial K(\vec{v})}{\partial\alpha_{j}}\right)\frac{\partial}{\partial\alpha_{j}^{+}}\,\right]\,\right\} \widehat{\Lambda}\, d^{4M+2}\vec{v},\notag\end{eqnarray}
with \begin{equation}
N_{j}=\alpha_{j}^{+}\alpha_{j},\end{equation}
which is initially the number of particles at the $j$-th site, and
an effective complex-variable Gibbs factor $K$ corresponding to $\mathrm{Tr}\left[{(\widehat{H}-\mu_{e}\widehat{N})\widehat{\Lambda}}\right]/\mathrm{Tr}\left[{\widehat{\Lambda}}\right]$:
\begin{equation}
K(\vec{v})=\sum_{j}\left\{ \frac{\hbar^{2}\left(\nabla\alpha_{j}^{+}\right)\left(\nabla\alpha_{j}\right)}{2m}-\mu_{e}N_{j}+\frac{gN_{j}^{2}}{2\Delta x}\right\} .\label{GibbsK}\end{equation}
Here $\nabla\alpha_{j}$ is the discretized analogue of the gradient
of a complex field $\alpha(x)$ that satisfies $\alpha(x_{j})=\alpha_{j}$.

To obtain a FPE equation for $G(\vec{v})$ we proceed as follows.
Firstly, we can make use of the additional {}``gauge\textquotedblright\
identity that follows trivially from Eq. (\ref{lambda}), \begin{equation}
\left(\Omega\frac{\partial}{\partial\Omega}-1\right)\widehat{\Lambda}=0,\label{gaugeidentity}\end{equation}
to convert $K(\vec{v})\widehat{\Lambda}=K(\vec{v})\Omega\frac{\partial}{\partial\Omega}\widehat{\Lambda}$
on the first line of Eq. (\ref{differential}). This step is necessary
in order to obtain an equation of a form that can later be sampled
with a diffusive process. Secondly, we integrate by parts to obtain
differentials of $G$ rather than $\widehat{\Lambda}$. Thirdly, if
the distribution $G$ is well bounded as $|\alpha_{j}|,|\alpha_{j}^{+}|,|\Omega|\rightarrow\infty$,
we can discard the boundary terms. As it turns out (see appendix~\ref{append:gauge}),
this is not fully justified for the equation (\ref{differential}),
and the boundary behavior will need to be improved with the help of
a stochastic gauge as described originally in \cite{deuar-drummond-2002}.
However, for demonstrative purposes let us proceed on for now, and
return to remedy the problem below in Sec.~\ref{sect:ito}. Lastly,
having now an equation of the form $\int\widehat{\Lambda}\times\lbrack\text{Differential operator}]G(\vec{v})\, d\vec{v}=0$,
one solution is certainly $[\text{Differential operator}]G(\vec{v})=0$,
which is the following FPE: \begin{widetext} \begin{equation}
0 = \left\{ \frac{\partial}{\partial\Omega}\Omega K(\vec{v})-\frac{\partial}{\partial\beta}-\sum_{j,\nu}\left[\frac{g}{4\Delta x}\frac{\partial^{2}}{\partial(\alpha_{j}^{(\nu)})^{2}}(\alpha_{j}^{(\nu)})^{2}+\frac{1}{2}\frac{\partial}{\partial\alpha_{j}^{(\nu)}}\left(\frac{\hbar^{2}(\nabla^{2}\alpha_{j}^{(\nu)})}{2m}+\mu_{e}\alpha_{j}^{(\nu)}-\frac{g\alpha_{j}^{(\nu)}N_{j}}{\Delta x}\right)\right]\right\} G(\vec{v}).\label{ppfpe}\end{equation}
 \end{widetext}

\subsection{Equivalent diffusion}

A diffusive random walk that corresponds to the Fokker-Planck equation
(\ref{ppfpe}) is found by replacing the analytic derivatives with
appropriate derivatives of the real and imaginary parts of $\alpha_{j}^{(\nu)}$
\cite{positiveP,Gardiner}. This results in a diffusion matrix in
the phase-space variables $\vec{v}$ with no negative eigenvalues.
In the Ito calculus this is equivalent to the following set of stochastic
differential equations \begin{eqnarray}
\frac{d\alpha_{j}^{(\nu)}}{d\beta} & = & \frac{1}{2}\left(\mu_{e}+\frac{\hbar^{2}\nabla^{2}}{2m}-\frac{gN_{j}}{\Delta x}\right)\alpha_{j}^{(\nu)}\notag\\
 &  & +i\alpha_{j}^{(\nu)}\sqrt{\frac{g}{2\Delta x}}\zeta_{j}^{(\nu)}(\beta),\label{ppequations}\\
\frac{d\Omega}{d\beta} & = & -\Omega K(\vec{v}).\notag\end{eqnarray}
We do not use diffusion gauges \cite{deuar-drummond-2006} here and
decompose the diffusion matrix in the most straightforward fashion.
Here, the $\zeta_{j}^{(\nu)}(\beta)$ are real, delta-correlated,
independent white Gaussian noise fields that satisfy the stochastic
averages \begin{subequations} \label{noises} \begin{eqnarray}
\langle\zeta_{j}^{(\nu)}(\beta)\rangle_{\mathcal{S}} & = & 0,\\
\langle\zeta_{i}^{(\nu)}(\beta)\zeta_{j}^{(\nu^{\prime})}(\beta^{\prime})\rangle_{\mathcal{S}} & = & \delta_{ij}\delta_{\nu\nu^{\prime}}\delta(\beta-\beta^{\prime}).\end{eqnarray}
In practice, at each time step separated from the subsequent by an
interval $\Delta\beta$, one generates $M$ independent real Gaussian
random variables of variance $1/\Delta\beta$ for each $\zeta_{j}^{(\nu)}$.

Equations (\ref{ppequations}) can be intuitively interpreted by noting
that the equation for the amplitudes $\alpha_{j}^{(\nu)}$ at each
point is a Gross-Pitaevskii equation in imaginary time, with some
extra noises that emulate the wandering of trajectories in a path
integral formulation around the mean field solution given by the deterministic
part. A different wander for different $\nu$. The weight evolution
of $\Omega$ generates the Gibbs factors of the grand canonical ensemble.

\subsection{Final equations}

\label{sect:ito}

A straightforward application of the diffusion equations (\ref{ppequations})
is foiled by the presence of an instability in the $d\alpha_{j}^{(\nu)}/d\beta$
equations. We use a stochastic gauge to remove this instability, in
a manner described in \cite{deuar-drummond-2006,deuar-thesis}, with
the details given in Appendix~\ref{append:gauge}. The final Ito
stochastic equations of the samples are \end{subequations} \begin{gather}
\frac{d\alpha_{j}^{(\nu)}}{d\beta}=\frac{1}{2}\left[\mu_{e}+\frac{\hbar^{2}\nabla^{2}}{2m}-\left(\frac{g}{\Delta x}\right)\left(|N_{j}|-i\,\text{Im}N_{j}\right)\right.\notag\\
\left.\hfill+i\zeta_{j}^{(\nu)}(\beta)\sqrt{\frac{2g}{\Delta x}}\,\right]\alpha_{j}^{(\nu)},\label{Gequations}\\
\frac{d\Omega}{d\beta}=\Omega\left[-K(\vec{v})-i\sqrt{\frac{g}{2\Delta x}}\sum_{j,\nu}\zeta_{j}^{(\nu)}(\beta)\left(|N_{j}|-\text{Re}N_{j}\right)\right].\notag\end{gather}
Some technical details regarding integration procedure, importance
sampling, and choice of $\mu_{e}(\beta)$ are given in Appendix~\ref{append:numerix}.
Attention to these issues can speed up the calculations and reduce
sampling errors by orders of magnitude.

\subsection{Evaluating observables}

Given $\mathcal{S}$ realizations of the variable sets $\vec{v}$,
using fresh initial samples and noises $\zeta_{j}^{(\nu)}(\beta)$
each time, one generates an estimate of the expectation value of an
observable $\widehat{O}$ as follows: \begin{eqnarray}
E\left[\widehat{O}\right]=\frac{\mathrm{Tr}\left[{\widehat{O}\widehat{\rho}_{u}}\right]}{\mathrm{Tr}\left[{\widehat{\rho}_{u}}\right]} & = & \frac{\int G(\vec{v})\mathrm{Tr}\left[{\widehat{O}\widehat{\Lambda}(\vec{v})}\right]\, d\vec{v}}{\int G(\vec{v})\mathrm{Tr}\left[{\widehat{\Lambda}(\vec{v})}\right]\, d\vec{v}}\notag\\
=\frac{\left\langle \mathrm{Tr}\left[{\widehat{O}\widehat{\Lambda}(\vec{v})}\right]\right\rangle _{\mathcal{S}}}{\left\langle \mathrm{Tr}\left[{\widehat{\Lambda}(\vec{v})}\right]\right\rangle _{\mathcal{S}}} & = & \frac{\text{Re}\left\langle {\mathcal{F}\left[\widehat{O},\vec{v}\right]}\right\rangle _{\mathcal{S}}}{\text{Re}\left\langle {\Omega}\right\rangle _{\mathcal{S}}},\label{obs}\end{eqnarray}
where $\langle\cdots\rangle_{\mathcal{S}}$ denotes a stochastic average
over the samples, and $\mathcal{F}$ is an appropriate function of
the phase-space variables $\vec{v}$. The last line follows from properties
of the operator basis $\widehat{\Lambda}$, and because the trace
of $\widehat{\rho}_{u}$ and of expectation values are real.

The identities (\ref{identities}) can be used to readily evaluate
$\mathcal{F}$ since $\mathrm{Tr}\left[{\widehat{\Lambda}}\right]=\Omega$.
In particular, \begin{equation}
\left\langle \widehat{\Psi}^{\dagger}(x_{j})\widehat{\Psi}(x_{j})\right\rangle =\frac{\text{Re}\left\langle ({N_{j}\Omega})\right\rangle _{\mathcal{S}}}{\Delta x\ \text{Re}\left\langle \Omega\right\rangle _{\mathcal{S}}},\label{expn}\end{equation}
\begin{equation}
\left\langle \widehat{\Psi}^{\dagger}(x_{i})\widehat{\Psi}^{\dagger}(x_{j})\widehat{\Psi}(x_{j})\widehat{\Psi}(x_{i})\right\rangle =\frac{\text{Re}\left\langle ({N_{i}N_{j}\Omega)}\right\rangle _{\mathcal{S}}}{(\Delta x)^{2}\text{Re}\left\langle {\Omega}\right\rangle _{\mathcal{S}}},\label{expnn}\end{equation}
which explains the relationship between $N_{j}$ and the particle
number at the $j$-th site. For the uniform system considered here,
it is efficient to average the quantities over the entire lattice,
so that e.g. \begin{equation}
g^{(2)}(r)=\frac{L\left\langle \int\widehat{\Psi}^{\dagger}(x)\widehat{\Psi}^{\dagger}(x+r)\widehat{\Psi}(x+r)\widehat{\Psi}(x)\, dx\right\rangle }{\left\langle \int\widehat{\Psi}^{\dagger}(x)\widehat{\Psi}(x)\, dx\right\rangle ^{2}}.\label{g2obs}\end{equation}

Uncertainty is estimated as follows: We separate the $\mathcal{S}$
realizations into $\mathcal{B}$ bins, such that $\mathcal{B}\gg1$
and $\mathcal{S}/\mathcal{B}\gg1$. One calculates an estimate for
the expectation value of an observable in each bin independently (let
us denote $\overline{O}_{i}$ as the estimate obtained from the $i$th
bin). The best estimate for the expectation value of the observable
is obviously $\langle\overline{O}_{i}\rangle_{\mathcal{B}}$. The
one-sigma uncertainty in this estimate is obtained from the Central
Limit theorem and is \begin{equation}
\Delta\overline{O}=\sqrt{\frac{\langle\overline{O}^{2}\rangle_{\mathcal{B}}-\langle\overline{O}\rangle_{\mathcal{B}}^{2}}{\mathcal{B}}}.\label{uncertainty}\end{equation}

\section{Nearly ideal gas regime {[}$\gamma\ll\min\{\tau^{2},\sqrt{\tau}\}$]}

\label{sect:nearlyideal}

We now present the perturbation theory results for the decoherent
regime of a 1D Bose gas \cite{karenprl}, where both the density
and phase fluctuations are large and the local pair correlation $g^{(2)}(0)$
is always close to the result for non-interacting bosons, $g^{(2)}(0)=2$.
Depending on the value of the temperature parameter $\tau$, we further
distinguish two sub-regimes: decoherent classical (DC) regime for
$\tau\gg1$ and decoherent quantum (DQ) regime for temperatures well
below quantum degeneracy, $\tau\ll1$. Both can be treated using perturbation
theory with respect to the coupling constant $g$ around the ideal
Bose gas, for which the nonlocal pair correlation function has been
studied in Ref.~\cite{Bouchoule-density-density}. Here, we extend
these results to account for the first-order perturbative terms.

\subsection{Perturbation theory in $\gamma$}

\label{sect:perturbation}

{ The correlations of a 1D Bose gas are governed by the action \begin{equation}
S\left[\Psi^{\ast}\Psi\right]=\int_{0}^{\beta}\! d\sigma\int\! dr\;\left[\Psi^{\ast}\partial_{\sigma}\Psi-{\cal H}(\Psi^{\ast},\Psi)\right],\label{eq:action}\end{equation}
 written in terms of a space and imaginary time dependent \textit{c}-number
fields $\Psi(x,\sigma)$ in the Feynman path integral formalism. Here
$\sigma$ is the imaginary time and $\beta=1/k_{B}T$ is the maximum,
corresponding to the inverse temperature. The Hamiltonian density
${\cal H}$ is obtained from (\ref{Hfull}) by replacing the operators
with the $c$-number fields. Using action (\ref{eq:action}), the
pair correlation function is given by \begin{equation}
g^{(2)}(r)=\frac{1}{n^{2}Z}\int\mathcal{D}\Psi^{\ast}\Psi\; e^{-S\left[\Psi^{\ast}\Psi\right]}\Psi^{\ast}(0)\Psi^{\ast}(r)\Psi(r)\Psi(0).\label{eq:g2def}\end{equation}
where $Z=\int\mathcal{D}\Psi^{\ast}\Psi\; e^{-S\left[\Psi^{\ast}\Psi\right]}$
is the partition function. In Eq.~(\ref{eq:g2def}) and below, we
use the notation that fields with imaginary time dependence omitted
act at $\sigma=0$, i.e. $\Psi(r)\equiv\Psi(r,0)$. 
Expanding the action (\ref{eq:action}) in powers of $g$, we obtain
up to the first order \begin{align}
g^{(2)}(r)= & g_{\mathrm{ideal}}^{(2)}(r)-\frac{g}{2n^{2}}\int_{0}^{\beta}\! d\sigma\int\! dr^{\prime}\;\langle\Psi^{\ast}(r^{\prime},\sigma)\Psi^{\ast}(r^{\prime},\sigma)\notag\\
 & \times\Psi(r^{\prime},\sigma)\Psi(r^{\prime},\sigma)\Psi^{\ast}(0)\Psi^{\ast}(r)\Psi(r)\Psi(0)\rangle,\label{expansion}\end{align}
where $g_{\mathrm{ideal}}^{(2)}(r)=1+G(r,0^{-})G(-r,0^{-})/n^{2}$ is
the ideal Bose gas result following from Wick's theorem. Note that
since the expansion above is formally in powers of $g$, the final
result can always be expressed in powers of $\gamma$ as
$\gamma\propto g$. The average in Eq.~(\ref{expansion}) is evaluated
using Wick's theorem \cite{note-Wick}
\begin{align}
\Delta g^{(2)}(r) & =g^{(2)}(r)-g_{\mathrm{ideal}}^{(2)}(r)=-\frac{2g}{n^{2}}\int_{0}^{\beta}\! d\sigma\int\! dr^{\prime}\;\label{eq:wick}\\
 & \times G(r^{\prime},\sigma)G(r-r^{\prime},-\sigma)G(r^{\prime}-r,\sigma)G(-r^{\prime},-\sigma),\notag\end{align}
with the Green's function \begin{eqnarray}
G(r,\sigma) & = & -\langle\Psi(0,0)\Psi^{\ast}(r,\sigma)\rangle\notag\\
 & = & \frac{1}{\beta L}\sum_{k,n}\frac{e^{ikr-i\hbar\omega_{n}\sigma}}{i\hbar\omega_{n}-\hbar^{2}k^{2}/2m+\mu}.\label{Grsimga}\end{eqnarray}
 The $\omega_{n}(\beta)$ are the Matsubara frequencies and the imaginary
time $\sigma$ runs between 0 and $\beta$. The Green's function is
periodic in the case of bosons and anti-periodic in the case of fermions.
Thus it can be Fourier transformed with $\omega_{n}=2\pi n/\beta$
(bosons) or $\omega_{n}=\pi(2n+1)/\beta$ (fermions). The discrete
sum over $k$ becomes an integral in thermodynamic limit. } 

In terms of a Green's
function $G_{k}(\sigma)$ that is Fourier transformed with respect
to the spatial coordinates, $\Delta g^{(2)}(r)$ can be brought to
the form \begin{equation}
\Delta g^{(2)}(r)=-\frac{2g}{n^{2}}\int_{0}^{\beta}\! d\sigma\!\int\frac{dk}{2\pi}e^{ikr}\Gamma(k,\sigma)\Gamma(k,-\sigma),\label{eq:weak_first}\end{equation}
where \begin{equation}
\Gamma(k,\sigma)=\frac{1}{2\pi}\int dp\ G_{p+k}(\sigma)G_{p}(-\sigma),\label{Gamma-1}\end{equation}
and \begin{equation}
G_{k}(\sigma)=\left\{ \begin{array}{cc}
-n_{k}(\beta)e^{-\sigma(\hbar^{2}k^{2}/2m-\mu)}, & \sigma<0,\\
-[1+n_{k}(\beta)]e^{-\sigma(\hbar^{2}k^{2}/2m-\mu)}, & \sigma>0,\end{array}\right.\label{G-k}\end{equation}
with \begin{equation}
n_{k}(\beta)=\frac{1}{e^{(\hbar^{2}k^{2}/2m-\mu)\beta}-1}\label{eq:bedistribution}\end{equation}
being the standard bosonic occupation numbers.


\subsection{Decoherent classical regime}

For temperatures above quantum degeneracy, $\tau\gg1$, the chemical
potential is large and negative, so the bosonic occupation numbers
are small, $n_{k}(\beta)\ll1$, and can be approximated by the Boltzmann
distribution, $n_{k}(\beta)\simeq e^{-(\hbar^{2}k^{2}/2m-\mu)\beta}$.
Accordingly, the function $G_{k}(\sigma)$ in Eq.~(\ref{G-k}) becomes
a Gaussian \begin{equation}
G_{k}(\sigma)=\left\{ \begin{array}{cc}
-\exp[-(\hbar^{2}k^{2}/2m-\mu)(\sigma+\beta)], & \sigma<0,\\
-\exp[-(\hbar^{2}k^{2}/2m-\mu)\sigma], & \sigma>0,\end{array}\right.\end{equation}
and Eq. (\ref{Gamma-1}) is integrated to yield \begin{equation}
\Gamma(k,\sigma)=\Gamma(k,-\sigma)=ne^{-\sigma(\beta-\sigma)\hbar^{2}k^{2}/2m\beta}.\label{eq:gamma_k_result}\end{equation}
Here the mean density at a given temperature and chemical potential
is determined from $n=\frac{1}{2\pi}\int dk\ G_{k}(0^{-})=\sqrt{m/(2\pi\hbar^{2}\beta)}\, e^{\beta\mu}$.
Using Eq. (\ref{eq:gamma_k_result}), the correction (\ref{eq:weak_first})
to the pair correlation function is found as (see Appendix \ref{append:integrals-nearly-ideal})
\begin{equation}
\Delta g^{(2)}(r)=-\gamma\sqrt{\frac{2\pi}{\tau}}\;\mathrm{erfc}\left(\sqrt{\frac{\tau n^{2}r^{2}}{2}}\right),\label{eq:weak_g2_res}\end{equation}
 where $\mathrm{erfc}(x)$ is the complimentary error function.

\begin{figure}
\includegraphics*[width=8cm]{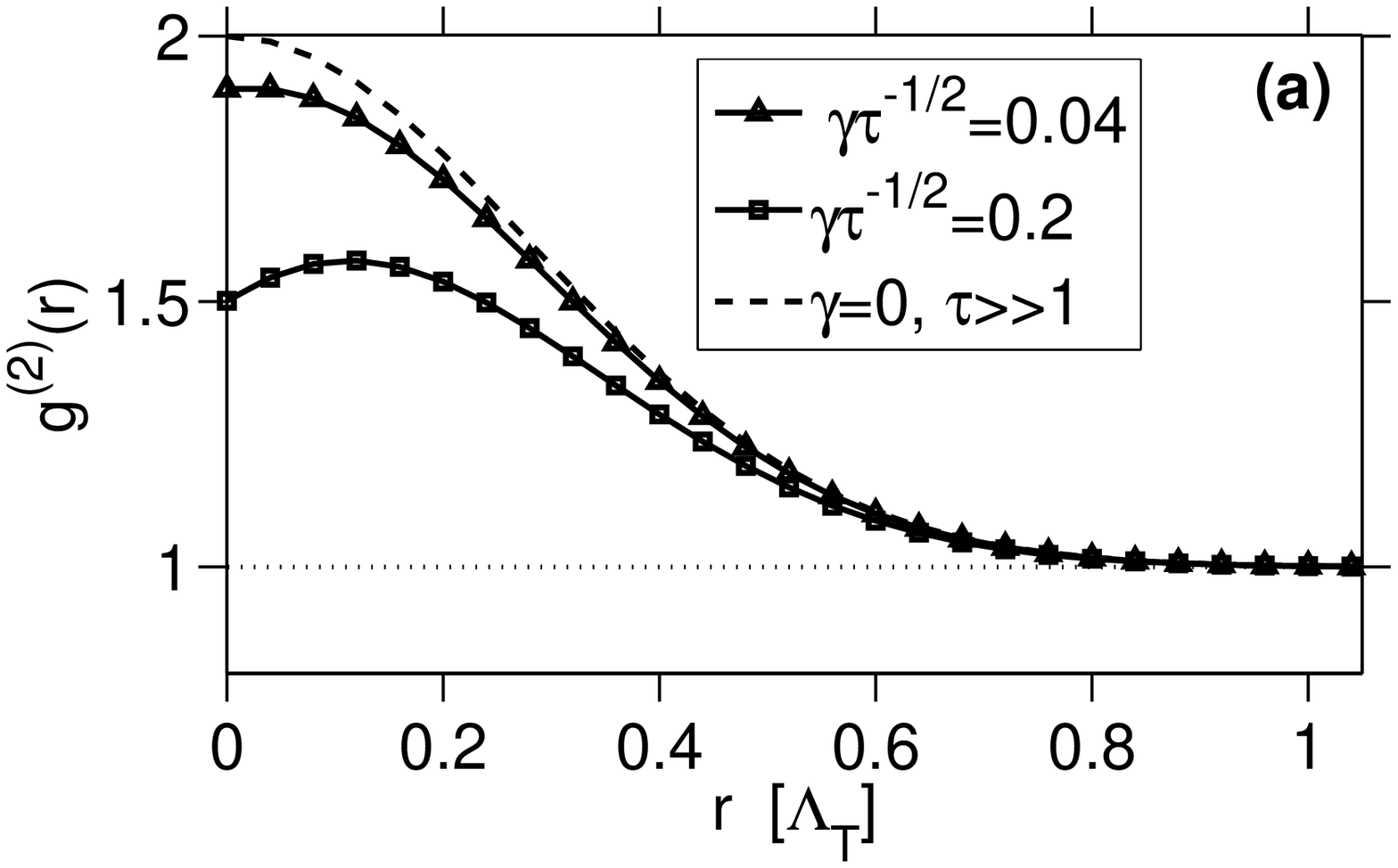}\\
\includegraphics*[width=8cm]{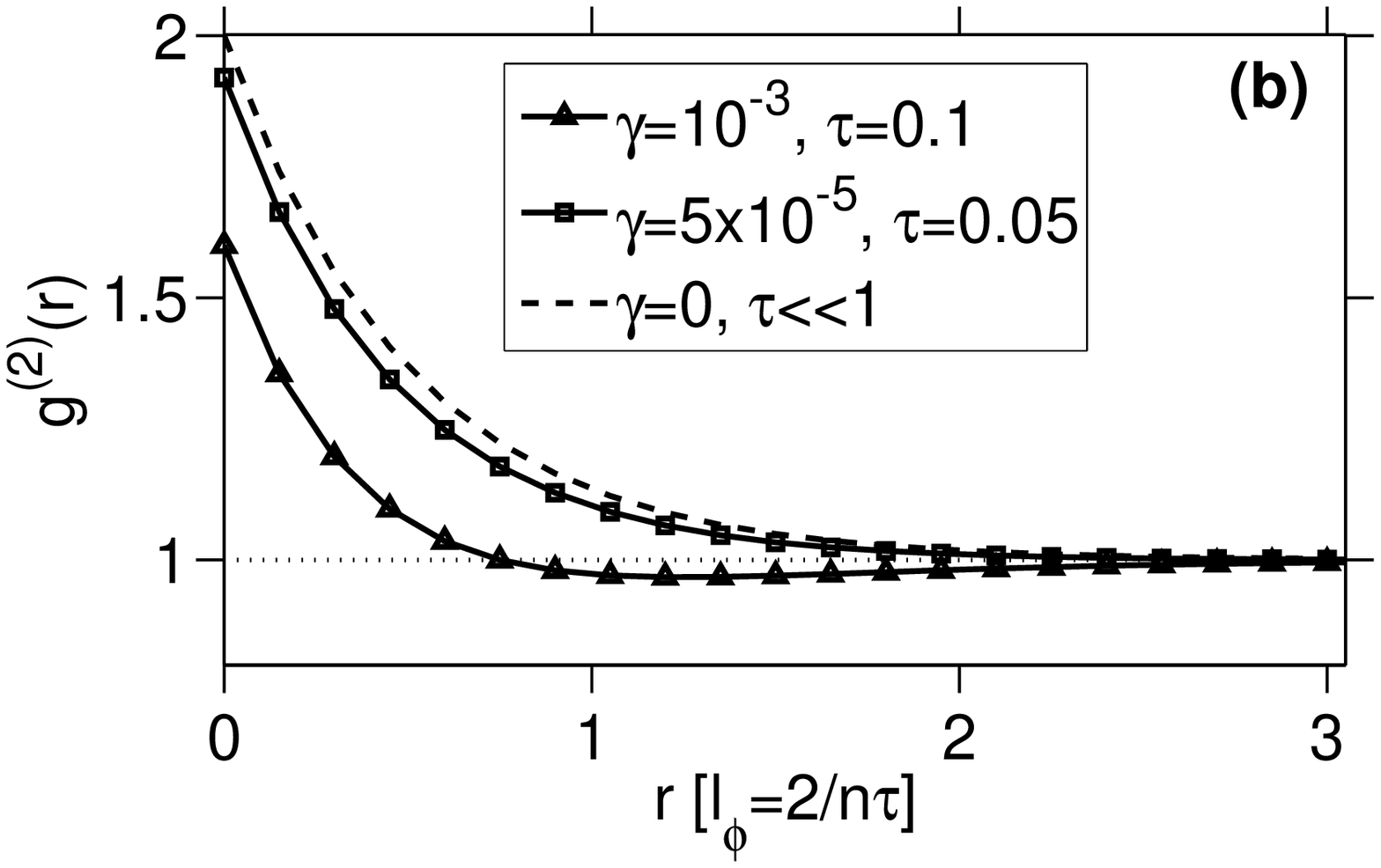}
\caption{Nonlocal pair correlation $g^{(2)}(r)$ in the nearly ideal
gas regime: (a) decoherent classical regime,
$\tau\gg\max\{1,\gamma^{2}\}$, Eq. (\protect\ref{DC}), with $r$ in
units of the thermal de Broglie wavelength
$\Lambda_{T}=\sqrt{4\pi/(\tau n^{2})}$; (b) decoherent quantum
regime, $\sqrt{\gamma}\ll\tau\ll1$, Eq. (\protect\ref{DQ}), with $r$
in units of the phase coherence length $l_{\phi}=2/n\tau$.}
\label{DQ_DC}
\end{figure}

Together with $g_{\mathrm{ideal}}^{(2)}(r)=1+\exp[-\tau n^{2}r^{2}/2]$
($\tau\gg1$), this gives the following result for the pair correlation
function in the DC regime ($\tau\gg\max\{1,\gamma^{2}\}$): \begin{equation}
g^{(2)}(r)=1+e^{-(r\sqrt{2\pi}/\Lambda_{T})^{2}}-\sqrt{\frac{2\pi\gamma^{2}}{\tau}}\;\mathrm{erfc}\left(\frac{r\sqrt{2\pi}}{\Lambda_{T}}\right),\label{DC}\end{equation}
This is written in terms of the thermal de Broglie wavelength \begin{equation}
\Lambda_{T}=\sqrt{\frac{2\pi\hbar^{2}}{2mT}}=\sqrt{\frac{4\pi}{\tau n^{2}}},\end{equation}
a quantity that will appear repeatedly in what follows. At $r=0$
we have $g^{(2)}(0)=2-\gamma\sqrt{2\pi/\tau}$ in agreement with Ref.
\cite{karenprl}. In the non-interacting limit ($\gamma=0$) we recover
the well-known result for the classical ideal gas \cite{Naraschewski-Glauber}
characterized by Gaussian decay with a correlation length $\Lambda_{T}$.
For $\gamma>0$ we observe {[}see Fig.~\ref{DQ_DC}(a)] the emergence
of anomalous behavior, with a global maximum $g^{(2)}(r_{\max})=g^{(2)}(0)+2\gamma^{2}/\tau$
at nonzero interparticle separation $nr_{\max}=2\gamma/\tau\ll1$.
{This corresponds to the emergence of antibunching, $g^{(2)}(0)<g^{(2)}(r_{\max})$,
due to repulsive interactions.} As $\gamma$ is increased further,
there is a continuous transition from the DC regime to the regime
of high-temperature {}``fermionization\textquotedblright\ (see
Sec. \ref{sect:fermionization}), with $g^{(2)}(0)$ reducing further
and the maximum moving to larger distances.

\subsection{Decoherent quantum regime}

{ For temperatures below quantum degeneracy, with $\sqrt{\gamma}\ll\tau\ll1$,
only $\omega_{n}=0$ contributes to the Green's function \begin{equation}
G_{k}(\sigma)=-T[\hbar^{2}k^{2}/(2m)+|\mu|]^{-1},\label{eq:green_boson_deg}\end{equation}
which gives the relation between the density and the chemical potential
$n=T\sqrt{m/(2\hbar^{2}|\mu|)}$, $\mu=-|\mu|$. Performing the Fourier
transform of Eq.~(\ref{eq:green_boson_deg}) one obtains the one-particle
density matrix for the ideal gas \begin{eqnarray}
g_{\text{ideal}}^{(1)}(r)=\langle\hat{\Psi}^{\dagger}(0)\hat{\Psi}(r)\rangle/n=\exp(-r/l_{\phi}),\label{eq:g1}\end{eqnarray}
 which characterizes the decay of phase coherence over a length scale
given by \begin{equation}
l_{\phi}=\frac{\hbar^{2}}{2m|\mu|}=\frac{2}{n\tau},\end{equation}
 and also determines the second-order correlation function for the
ideal gas \begin{equation}
g_{\text{ideal}}^{(2)}(r)=1+|g_{\text{ideal}}^{(1)}(r)|^{2}=1+e^{-2r/l_{\phi}}.\label{eq:g2ideal}\end{equation}
 }

{ The one-particle Greens function, Eq. (\ref{eq:green_boson_deg}),
together with Eq. (\ref{Gamma-1}) leads to $\Gamma(k,\sigma)=4n^{2}l_{\phi}/(k^{2}l_{\phi}^{2}+4)$.
Inserting it into Eq.~(\ref{eq:weak_first}) we obtain (see Appendix
\ref{append:integrals-nearly-ideal}) corrections to $g_{\text{ideal}}^{(2)}(r)$,
leading to the following result for the pair correlation function
in the DQ regime \begin{equation}
g^{(2)}(r)=1+\left[1-\frac{4\gamma}{\tau^{2}}\left(1+\frac{2r}{l_{\phi}}\right)\right]e^{-2r/l_{\phi}}.\label{DQ}\end{equation}
This has the maximum value $g^{(2)}(0)=2-4\gamma/\tau^{2}$, in agreement
with the result of Ref. \cite{karenprl}. For $\gamma=0$ the correlations
decay exponentially with the characteristic correlation length of
half a phase coherence length describing the long-wavelength phase
fluctuations. }

An interesting feature in this regime is the apparent prediction of
weak antibunching \emph{at a distance} as seen in Fig.~\ref{DQ_DC}
(b), with $g^{(2)}(r_{\min})<1$. The strongest antibunching in expression
(\ref{DQ}) occurs at $nr_{\min}=\tau/4\gamma\gg1$, or $r_{\min}=l_{\phi}\tau^{2}/4\gamma\gg l_{\phi}$,
and dips below unity by an amount $(4\gamma/\tau^{2})\exp(-\tau^{2}/4\gamma)\ll1$.
However, there is ambiguity regarding its existence: One should note
that the dip below unity is very small in the region of uncontested
validity of Eq.~(\ref{DQ}) where $\tau/\sqrt{\gamma}\gg1$, and
only becomes appreciable around $\tau\lesssim2\sqrt{\gamma}$, which
is in the crossover region into the quasi-condensate (see Sec.~\ref{sect:weak}).
Whether such anomalous antibunching survives higher order corrections
in the small parameter $\sqrt{\gamma}/\tau$ remains to be seen. Our
numerical calculations to date have not been able to access a regime
of small enough $\sqrt{\gamma}/\tau$ to confirm or deny its existence.

The numerical examples shown in Fig.~\ref{DQ_lowgam} are for $\sqrt{\gamma}/\tau\simeq0.24$
and $\sqrt{\gamma}/\tau\simeq0.77$, and show a thermal bunching peak
with a typical Gaussian shape at the shortest range of $\Lambda_{T}$,
with $\Lambda_{T}\ll l_{\phi}$. At longer ranges, phase coherence
dominates this and leads to exponential decay on the length scale
$l_{\phi}$, in agreement with Eq.~(\ref{DQ}).

\subsection{Quantum/classical transition}

\label{subsect:DC/DQ} 
\begin{figure}
\includegraphics*[width=8cm]{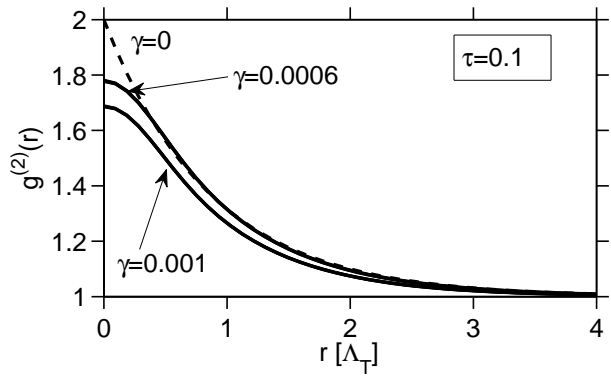}
\caption{Approach of the pair correlation function to the ideal gas
solution (shown dashed) in the decoherent quantum regime at
$\tau=0.1$, with $r$ in units of the thermal de Broglie wavelength,
$\Lambda_{T}=\sqrt{4\pi/\tau n^{2}}$. The thickness of the solid
lines (numerical results) comes from the superimposed $1\sigma$
error bars which are below resolution.}
\label{DQ_lowgam}
\end{figure}

\begin{figure}
\includegraphics*[width=8cm]{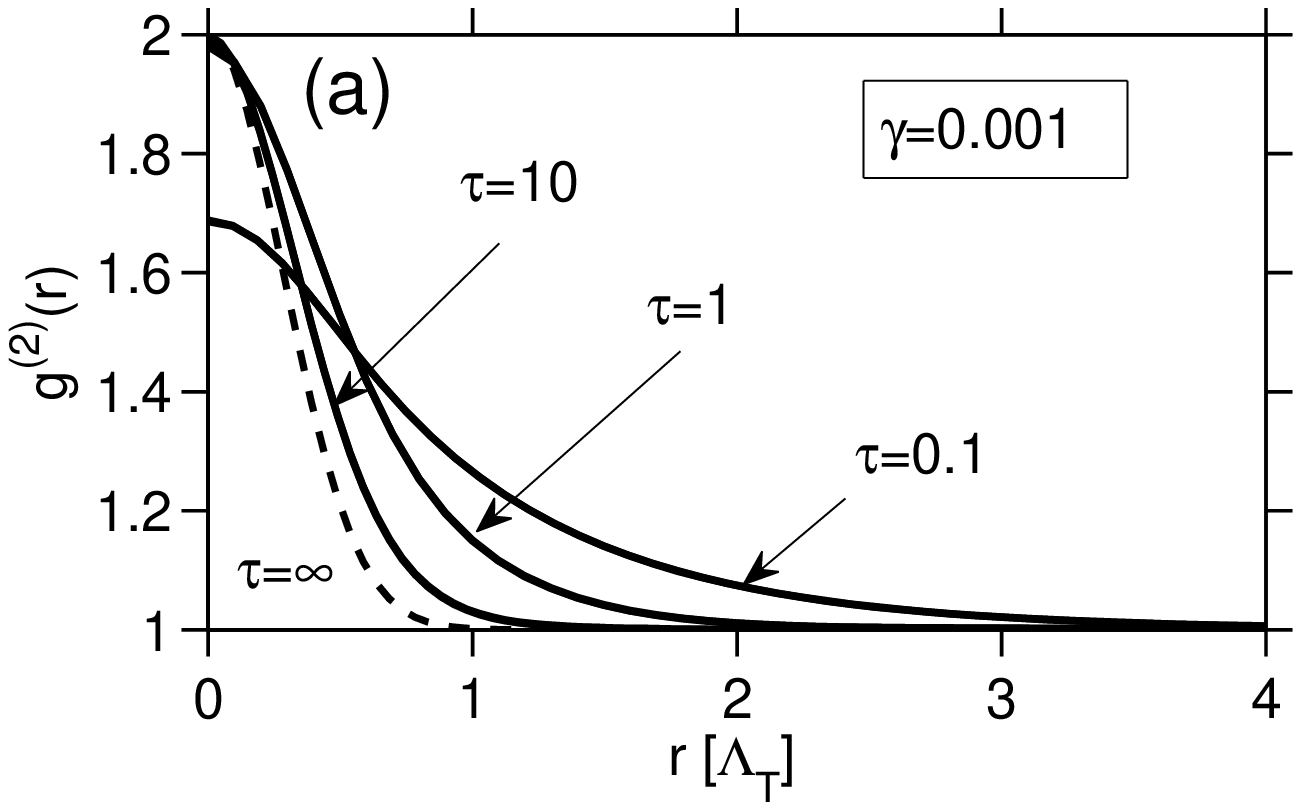}\\
\includegraphics*[width=8cm]{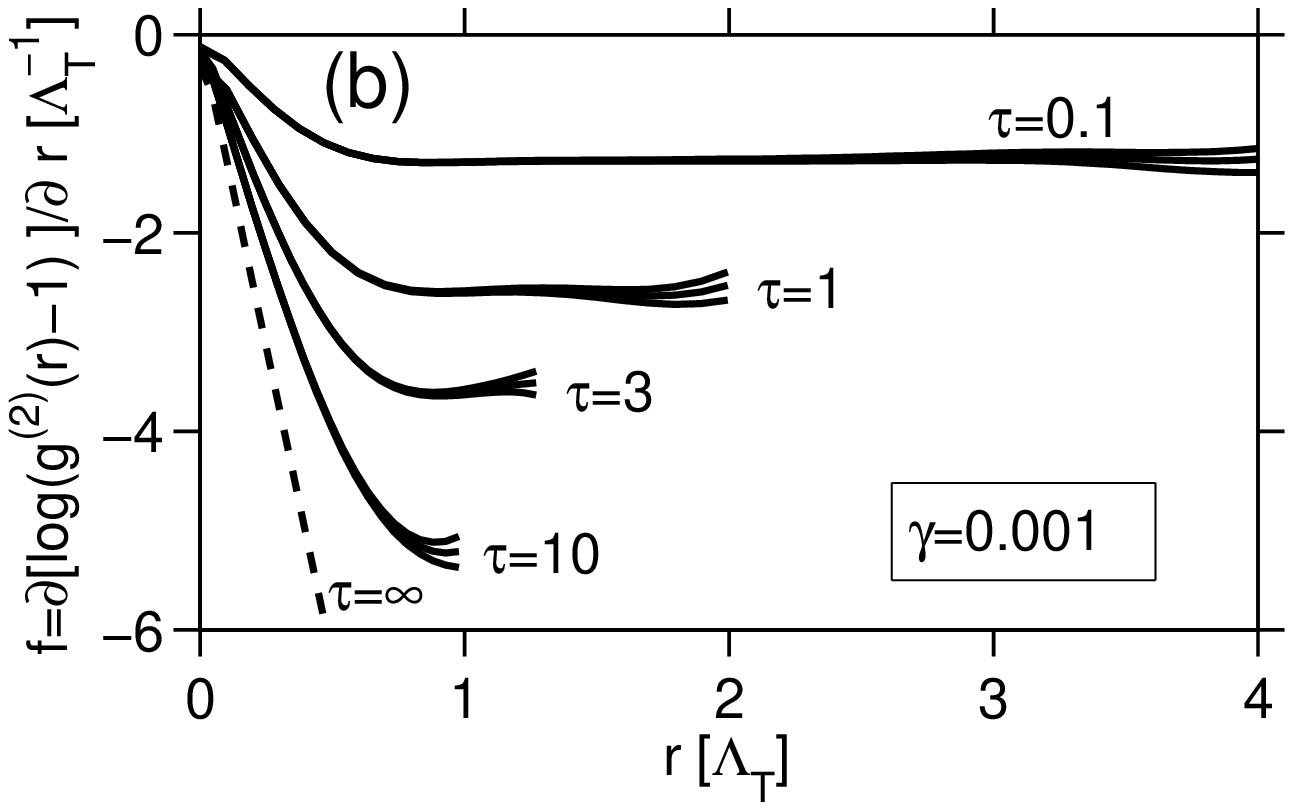}
\caption{Exact behavior of $g^{(2)}(r)$, with $r$ in units of
$\Lambda_{T}$, in the nearly ideal gas regime with $\gamma=0.001$
and varying $\tau$ around the quantum/classical crossover. In panel
(b), the derivative $f=\partial\lbrack\ln(g^{(2)}(r)-1)]/\partial r$
shows a clear distinction between exponential decay (when $f$ is
constant) and Gaussian thermal-like behavior when $f$ is linear. The
triple lines indicate the numerical curves together with $1\sigma$
error bars which are mostly below resolution.}

\label{lowgam}
\end{figure}

\begin{figure}
 \includegraphics*[width=8cm]{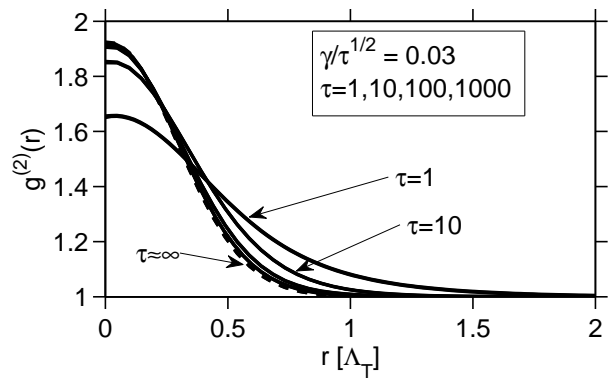}

\caption{Approach to the classical decoherent gas solution (shown dashed),
Eq. (\protect\ref{DC}), for finite but small interaction with $\gamma/\sqrt{\tau}=0.03$,
which corresponds to a variation of density while keeping the coupling
$g$ and $T$ constant. Here $g^{(2)}(0)\rightarrow1.925$ in the
$\tau\rightarrow\infty$ or equivalently $n\rightarrow0$ limit. Triple
solid lines are the numerical results, with $1\sigma$ error bars
below resolution.}

\label{DC_constn}
\end{figure}

\begin{figure}
 \includegraphics*[width=8cm]{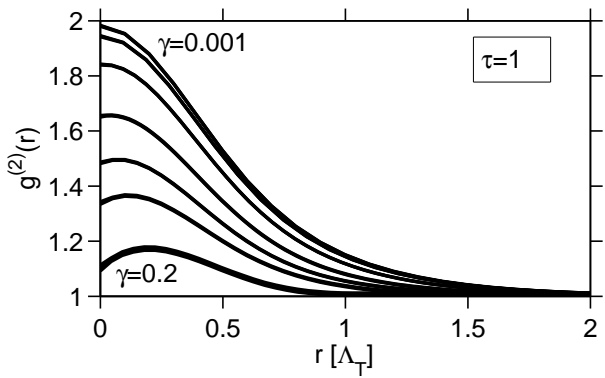}

\caption{Behavior of $g^{(2)}(r)$ in the crossover region between decoherent
classical and quantum gas at $\tau=1$. Values of $\gamma$ shown
are $0.001$, $0.003$, $0.01$, $0.03$, $0.06$, $0.1$ and $0.2$
as the curves for $g^{(2)}(r)$ descend. }

\label{tau1}
\end{figure}

The transition from the quantum to the classical
decoherent gas was investigated using the gauge-$P$ numerical method.
The behavior is shown in Figs.~\ref{DQ_lowgam}--\ref{tau1}.

With rising temperature, still below degeneracy, one first finds a
rounding-off of the exponential behavior at short ranges of a fraction
of $\Lambda_{T}$, as seen in Fig.~\ref{DQ_lowgam}. There is also
a global lowering of $g^{(2)}(r)$ with $\gamma$. It should be noted
that the parameters for the numerical results shown in Fig.~\ref{DQ_lowgam}
are not deep in the regime where (\ref{DQ}) applies accurately, and
the lowering of the tails with $\gamma$ is weaker here, than predicted
by that limiting expression.

Considering variation with $T$, as temperature approaches, and then
exceeds $T_{d}$, Gaussian thermal-like behavior appears first at
short ranges, progressively taking over an ever larger part of $g^{(2)}(r)$
as temperature is raised. This is seen in Fig.~\ref{lowgam}. The
exponential tails can persist at ranges $r\gtrsim\Lambda_{T}/\sqrt{2\pi}$
well into the high temperature regime when $\gamma$ is small, as
seen in Fig.~\ref{lowgam}(b) for $\tau=3$ and even $\tau=10$.

There are three scenarios that can typically be controlled in ultracold
gas experiments: (\textit{i}) varying the absolute temperature changes
$\tau$ but not $\gamma$, as in Fig.~\ref{lowgam}; (\textit{ii})
varying the coupling strength via a Feshbach resonance or varying
the width of the trapping potential affects $\gamma$ but not $\tau$,
as considered in Section~\ref{sect:HTFxover} and Fig.~\ref{DQ_lowgam};
and (\textit{iii)} varying the linear density gives changes in both
$\gamma$ and $\tau$, while keeping the quantity $\gamma/\sqrt{\tau}$
constant. Notably, this is the parameter that appears in the analytic
expressions for both decoherent regimes, Eqs.~(\ref{DQ}) and (\ref{DC}).

Figure~\ref{DC_constn} shows the behavior under scenario (\textit{iii}),
where increasing $\tau$ corresponds to decreasing density of the
gas. As expected, $g^{(2)}(0)$ tends to a constant value $g^{(2)}(0)=2-\gamma\sqrt{2\pi/\tau}\neq2$
with $\tau\rightarrow\infty$ predicted by Eq. (\ref{DC}). Interestingly,
the crossover is quite broad under changing density, with departures
from the decoherent classical result still visible at $\tau\sim100$.

Finally, in the middle of the crossover region at $\tau=1$, $\gamma\ll1$,
there is the smooth and quite broad transition from low$\ $values
of $\gamma$ to $\gamma\sim{\mathcal{O}}(1)$ that is shown in Fig.~\ref{tau1}.
The situation of a short-range Gaussian with standard deviation $\sim\Lambda_{T}/2\sqrt{\pi}$
and exponential tails with length scale $l_{\phi}/2$ that was seen
in Fig.~\ref{lowgam} morphs into an anomalous form with a local
maximum that is similar to the high temperature fermionization behavior
described below in Sections~\ref{sect:strong} and \ref{sect:HTFxover}.

\section{Weakly interacting quasi-condensate regime {[}$\tau^{2}\ll\gamma\ll1$]}

\label{sect:weak}

In the regime of weak interactions and low temperature (or Gross-Pitaevskii
regime) with $\gamma\ll1$ we rely on the fact that the equilibrium
state of the gas is that of a quasi-condensate~\cite{mermin-wagner-hohenberg,petrov-1d-regimes}.
In this regime the density fluctuations are suppressed while the phase
still fluctuates. The pair correlation function is close to one and
the deviations can be calculated using the Bogoliubov theory. In this
approach, the field operator $\hat{\Psi}$ is represented as a sum
of the ($c$-number) macroscopic component $\Psi_{0}$, containing
excitations with momenta $k\lesssim k_{0}\ll\xi^{-1}$ (where $\xi=\hbar/\sqrt{mgn}$
is the healing length) and a small operator component $\delta\hat{\Psi}$
describing excitations with larger momenta, $\hat{\Psi}=\Psi_{0}+\delta\hat{\Psi}$.
The momentum $k_{0}$ is chosen such that most of the particles are
contained in $\Psi_{0}$, however, its details do not enter into the
lowest order corrections to $g^{(2)}(r)$, which are $\mathcal{O}(\delta\hat{\Psi})^{2}$.
Using Wick's theorem, and the property of the thermal density matrix
that $\langle\delta\hat{\Psi}\rangle=0$, the pair correlation function
is then reduced to\begin{equation}
g^{(2)}(r)\simeq1+\frac{2}{n}\left(\text{Re}\langle\delta\hat{\Psi}^{\dagger}(r)\delta\hat{\Psi}(0)\rangle+\text{Re}{\langle\delta\hat{\Psi}(r)\delta\hat{\Psi}(0)\rangle}\right).\label{g2-Bog}\end{equation}
 The normal and anomalous averages $\langle\delta\hat{\psi}^{\dagger}(r)\delta\hat{\psi}(0)\rangle$
and $\langle\delta\hat{\psi}(r)\delta\hat{\psi}(0)\rangle$ are calculated
using the Bogoliubov transformation\begin{equation}
\delta\hat{\psi}(r)=\frac{1}{L}\sum\nolimits _{k}\left(u_{k}\hat{a}_{k}e^{ikx}-v_{k}\hat{a}_{k}^{\dagger}e^{-ikx}\right),\end{equation}
where $L$ is the length of the quantization box, $\hat{a}_{k}$ and
$\hat{a}_{k}^{\dagger}$ are the annihilation and creation operators
of elementary excitations, and $(u_{k},v_{k})$ are the expansion
coefficients given by \begin{equation}
u_{k}=\frac{\epsilon_{k}+E_{k}}{2\sqrt{\epsilon_{k}E_{k}}},\; v_{k}=\frac{\epsilon_{k}-E_{k}}{2\sqrt{\epsilon_{k}E_{k}}},\label{u-v}\end{equation}
and satisfying $u_{k}^{2}-v_{k}^{2}=1$. Here $\epsilon_{k}=\sqrt{E_{k}(E_{k}+2gn)}$
is the Bogoliubov excitation energy, $E_{k}=\hbar^{2}k^{2}/(2m)$,
and we note that the following useful relationships between $E_{k}$
and $\epsilon_{k}$ hold:\begin{eqnarray}
E_{k} & = & \sqrt{\epsilon_{k}^{2}+(gn)^{2}}-gn,\\
\frac{E_{k}}{\epsilon_{k}} & = & \left[\frac{k^{2}}{k^{2}+(2/\xi)^{2}}\right]^{1/2},\label{E_over_eps}\end{eqnarray}
where $\xi=\hbar/\sqrt{mgn}$ is the healing length. The equilibrium
occupation numbers of the Bogoliubov excitations are given by $\tilde{n}_{k}=\langle\hat{a}_{k}^{\dagger}\hat{a}_{k}\rangle=[e^{\epsilon_{k}/T}-1]^{-1}$.

Applying the Bogoliubov transformation to the normal and anomalous
averages in Eq. (\ref{g2-Bog}) gives \begin{eqnarray}
g^{(2)}(r) & = & 1+\frac{1}{\pi n}\int\limits _{-\infty}^{+\infty}dk\,\cos(kr)\notag\\
 &  & \times\left[(u_{k}-v_{k})^{2}\tilde{n}_{k}+v_{k}(v_{k}-u_{k})\right].\end{eqnarray}

Using next Eq. (\ref{u-v}) for the coefficients $u_{k}$ and $v_{k}$
we obtain the following result for the pair correlation function \begin{equation}
g^{(2)}(r)=1+\frac{1}{2\pi n}\int\limits _{-\infty}^{+\infty}dk\left[\frac{E_{k}}{\epsilon_{k}}(2\tilde{n}_{k}+1)-1\right]\cos(kr).\end{equation}

For convenience, we split the $g^{(2)}(r)$-function into two parts
corresponding to the contributions of thermal and vacuum fluctuations,
\begin{equation}
g^{(2)}(r)=1+G_{0}(r)+G_{T}(r),\label{g2r-split}\end{equation}
with\begin{equation}
G_{0}(r)=\frac{1}{2\pi n}\int\limits _{-\infty}^{+\infty}dk\left[\frac{E_{k}}{\epsilon_{k}}-1\right]\cos(kr),\label{G-0-def}\end{equation}
and \begin{equation}
G_{T}(r)=\frac{1}{\pi n}\int\limits _{-\infty}^{+\infty}dk\frac{E_{k}}{\epsilon_{k}}\tilde{n}_{k}\cos(kr).\label{G-T-def}\end{equation}

We first evaluate the vacuum contribution $G_{0}(r)$, Eq. (\ref{G-0-def}).
As shown in Appendix \ref{append:Bogoliubov}, the integral in (\ref{G-0-def})
can be obtained exactly in terms of special functions, giving \begin{equation}
G_{0}(r)=-\sqrt{\gamma}\left[\mathbf{L}_{-1}(2\sqrt{\gamma}nr)-I_{1}(2\sqrt{\gamma}nr)\right],\label{G0-specialf}\end{equation}
where $\mathbf{L}_{-1}(x)$ is the modified Struve function and $I_{1}(x)$
is a Bessel function. {The correlation length scale here is set by
the healing length $\xi=\hbar/\sqrt{mgn}=1/\sqrt{\gamma}n$.}

\subsection{{Quasi-condensate at low temperatures}}

At very low temperatures when the excitations are dominated by vacuum
fluctuations, whereas the thermal fluctuations are a small correction,
the $G_{T}(r)$-term is calculated as follows. First, we substitute
the explicit expression for $\tilde{n}_{k}$ into Eq. (\ref{G-T-def}),
giving\begin{equation}
G_{T}(r)=\frac{1}{\pi n}\int\limits _{-\infty}^{+\infty}dk\frac{E_{k}}{\epsilon_{k}^{\,}}\frac{1}{e^{\epsilon_{k}/T}-1}\cos(kr).\label{G-T-lowT}\end{equation}
As shown in Appendix \ref{append:Bogoliubov}, for $T\ll gn$ (or
$\tau\ll\gamma$) the integral can be simplified and gives

\begin{equation}
G_{T}(r)\simeq\frac{\pi}{2\sqrt{\gamma}}\left[\frac{1}{n^{2}\pi^{2}r^{2}}-\frac{\tau^{2}}{4\gamma}\cosech^{2}\left(\frac{\pi\tau nr}{2\sqrt{\gamma}}\right)\right].\label{G-T-GPa}\end{equation}

\begin{figure}
\includegraphics*[width=8cm]{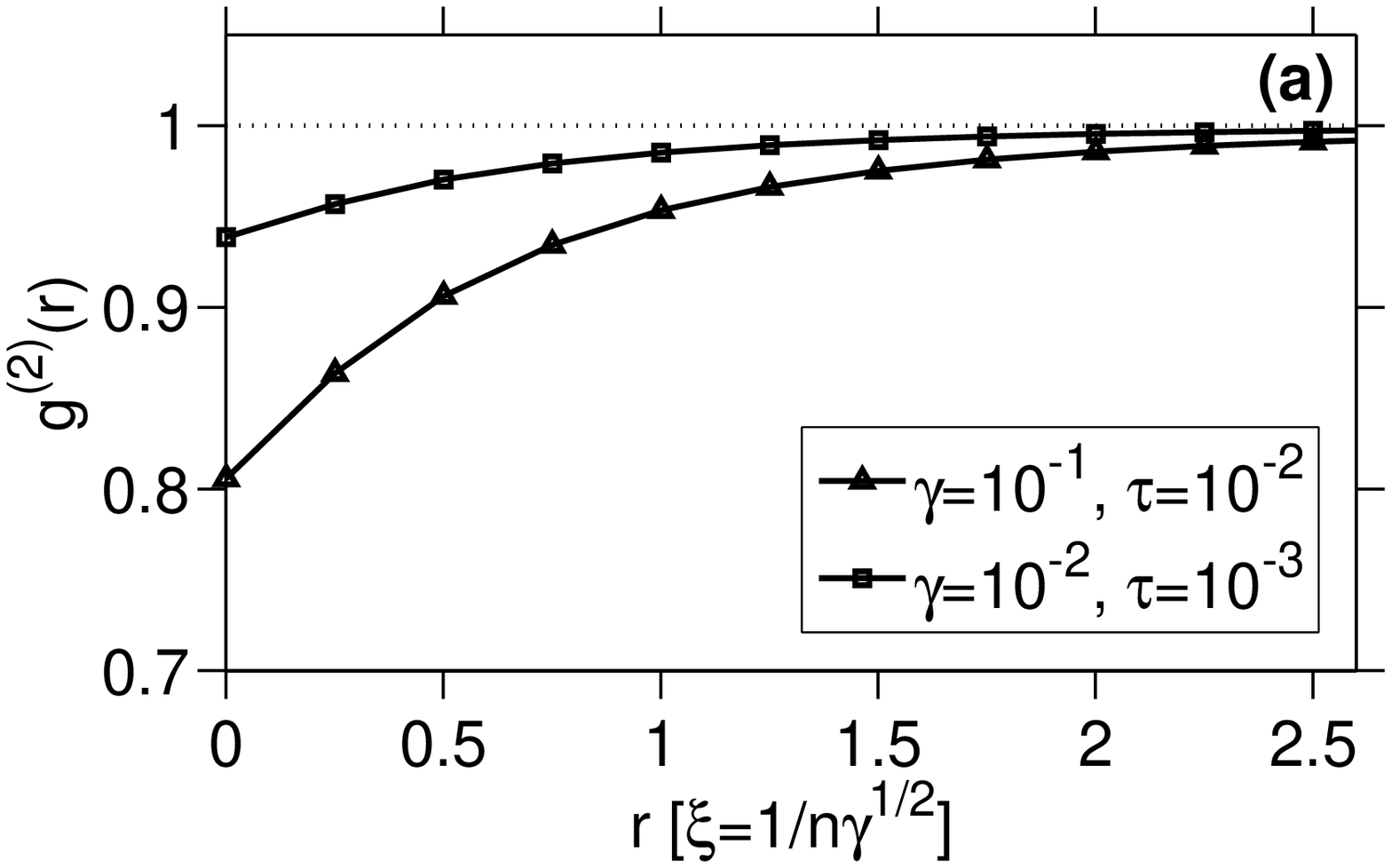}\\
\includegraphics*[width=8cm]{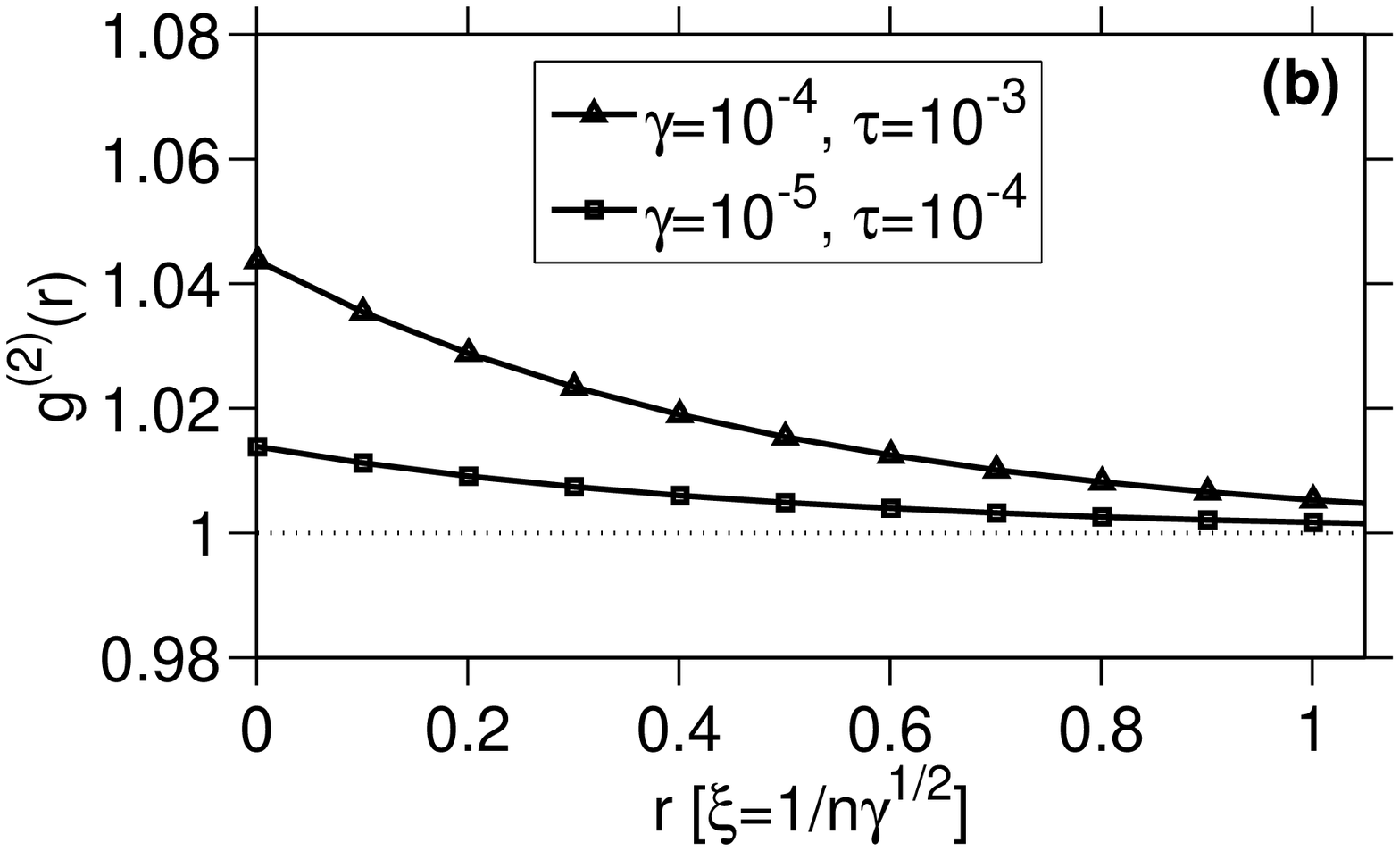}

\caption{Nonlocal pair correlation $g^{(2)}(r)$ in the weakly interacting
regime, with $r$ in units of the healing length $\xi=1/\sqrt{\gamma}n$:
(a) low-temperature weekly interacting gas at $\tau\ll\gamma\ll1$,
Eq. (\protect\ref{GPa}); (b) weakly interacting gas at $\gamma\ll\tau\ll\sqrt{\gamma}$,
Eq. (\protect\ref{GPb}).}

\label{GP}
\end{figure}

Combining Eqs. (\ref{g2r-split}), (\ref{G0-specialf}) and (\ref{G-T-GPa})
we obtain the following final result for this regime ($\tau\ll\gamma\ll1$):\begin{align}
g^{(2)}(r) & =1-\sqrt{\gamma}\left[\mathbf{L}_{-1}(2r/\xi)-I_{1}(2r/\xi)\right]\notag\\
 & +\frac{\sqrt{\gamma}\xi^{2}}{2\pi r^{2}}-\frac{\pi\tau^{2}}{8\gamma^{3/2}}\sinh^{-2}\left(\frac{\pi\tau r}{2\gamma\xi}\right).\label{GPa}\end{align}
{ In the limit of $\tau\rightarrow0$, the terms in the second line
of Eq.~(\ref{GPa}) cancel each other and the large distance ($r\gg\xi$)
asymptotics of the difference of special functions $\mathbf{L}_{-1}(x)-I_{1}(x)\sim1/8\pi x^{2}$
ensures the expected inverse square decay of correlations\cite{giamarchi-book}.
At small but finite temperatures, the same large-distance asymptotics
exactly cancels the inverse square behavior in the second line of
Eq.~(\ref{GPa}) leaving only the exponential decay \begin{eqnarray}
g^{(2)}(r)\underset{r\rightarrow\infty}{\longrightarrow}1-\frac{\pi\tau^{2}}{8\gamma^{3/2}}e^{-\pi\tau r/\gamma\xi}\label{eq:g2r_Bogo_exp_decay}\end{eqnarray}
 to the uncorrelated value of $g^{(2)}(r)=1$. This is again in full
agreement with the Luttinger liquid theory \cite{giamarchi-book}.
} We note that even at $T=0$, oscillating terms are absent, in contrast
to the strongly interacting regime of Sec. \ref{sect:strong-Tonks},
Eq.~(\ref{TGa}). The limit $r\rightarrow0$ in Eq.~(\ref{GPa})
reproduces the result of Eq.~(9) of Ref.~\cite{karenprl}, $g^{(2)}(0)=1-2\sqrt{\gamma}/\pi+\pi\tau^{2}/(24\gamma^{3/2})$.
In Fig.~\ref{GP}(a) we plot Eq.~(\ref{GPa}) for different values
of the interaction parameter $\gamma$, and we note that the finite
temperature correction term is negligible here.

\subsection{Thermally excited quasi-condensate}

In the opposite limit, dominated by thermal rather than vacuum fluctuations
and corresponding to $\gamma\ll\tau\ll\sqrt{\gamma}$, the thermal
part of the pair correlation function is calculated as follows. We
first note that large thermal fluctuations correspond to $\tilde{n}_{k}\gg1$,
which in turn requires $\epsilon_{k}/T\ll1$. Thus, we replace $\tilde{n}_{k}$
in the integral (\ref{G-T-def}) by $\tilde{n}_{k}=[\exp(\epsilon_{k}/T)-1]^{-1}\simeq T/\epsilon_{k}\gg1$.
{ With this substitution, the integral for $G_{T}(r)$ is dominated
by the free-particle (quadratic in $k$) part of the Bogoliubov spectrum
and the calculations in Appendix~\ref{append:Bogoliubov} yield \begin{equation}
G_{T}(r)=\frac{\tau}{2\sqrt{\gamma}}e^{-2\sqrt{\gamma}nr}.\label{G-T-GPb}\end{equation}
This result is valid for $r/\xi\lesssim1$. For $r/\xi\gg1$ the main
contribution to the integral in Eq.~(\ref{G-T-def}) comes from the
phonon (linear in $k$) part of the Bogoliubov spectrum and one recovers
the behavior given by Eq.~(\ref{eq:g2r_Bogo_exp_decay}). }

Combining Eqs.~(\ref{g2r-split}), (\ref{G0-specialf}) and (\ref{G-T-GPb})
we obtain the following final result for this regime ($\gamma\ll\tau\ll\sqrt{\gamma}$
{and $r\lesssim\xi$}):\begin{align}
g^{(2)}(r) & =1+\frac{\tau}{2\sqrt{\gamma}}e^{-2r/\xi}\notag\\
 & -\sqrt{\gamma}\left[\mathbf{L}_{-1}(2r/\xi)-I_{1}(2r/\xi)\right].\label{GPb}\end{align}
The last two terms are due to vacuum fluctuations and are a negligible
correction here, so the leading term gives an exponential decay of
correlations {[}see Fig.~\ref{GP}(b)] with a characteristic correlation
length given by the healing length $\xi=1/\sqrt{\gamma}n$. The peak
value at $r=0$ is $g^{(2)}(0)=1+\tau/(2\sqrt{\gamma})$, in agreement
with Ref.~\cite{karenprl}.

\section{Strongly interacting regime {[}$\gamma\gg\max\{1,\sqrt{\tau}\}$]}

\label{sect:strong}

\subsection{Perturbation theory in $1/\gamma$}

By mapping the system onto that of a weakly attractive 1D fermion
gas \cite{Cheon-Shigevare} one can perform perturbation theory in
$1/\gamma\ll1$. The formalism is the same as in Sec.
\ref{sect:perturbation}, except that $\Psi$ is now a fermionic field
and the interaction term in the Hamiltonian (\ref{Hfull}) has to be
modified to describe effective attractive interaction between
fermions with matrix elements (in $k$-space)
$V_{k}=-2\hbar^{2}k^{2}/(mn\gamma)$ \cite{Cheon-Shigevare}. Then
\[
g^{(2)}(r)=g_{\gamma=\infty}^{(2)}(r)+\Delta g^{(2)}(r)\]
with $g_{\gamma=\infty}^{(2)}(r)=1-e^{-n^{2}\tau r^{2}/2}$. The first
order corrections to $g^{(2)}(r)$ are given by the Hartree-Fock approximation
as a sum of the direct and exchange contributions 
\begin{align}
\Delta g_{d}^{(2)}(r) & =\int_{0}^{\beta}\! d\sigma\!\int\frac{dk}{2\pi}\; V_{k}\Gamma(k,\sigma,r=0)\Gamma(-k,\sigma,r=0)e^{ikr},\label{eq:the_diagram}\\
\Delta g_{e}^{(2)}(r) & =-\int_{0}^{\beta}\! d\sigma\!\int\frac{dk}{2\pi}\; V_{k}\Gamma(k,\sigma,r)\Gamma(-k,\sigma,-r)e^{ikr},\label{delta_g_exchange}\end{align}
where \begin{equation}
\Gamma(k,\sigma,r)=\int dp\ G_{p+k}(\sigma)G_{p}(-\sigma)e^{ipr}/2\pi,\label{Gamma-2}\end{equation}
in terms of the Green's function $G_{k}(\sigma)$ for free fermions.

\subsection{Regime of high-temperature {}``fermionization\textquotedblright}

\label{sect:fermionization}

We proceed with evaluation in the regime of high-temperature {}``fermionization\textquotedblright\ at
temperatures well above quantum degeneracy, $\tau\gg1$. In this regime,
we use the Maxwell-Boltzmann distribution of quasi-momenta as the
unperturbed state. In the temperature interval $1\ll\tau\ll\gamma^{2}$,
the characteristic distance related to the interaction between the
particles -- the 1D scattering length $a_{1D}=\hbar^{2}/mg\simeq l_{\perp}^{2}/a$
$\sim1/\gamma n$ -- is much smaller than the thermal de Broglie wavelength
$\Lambda_{T}$, and the small perturbation parameter is $a_{1D}/\Lambda_{T}\ll1$
\cite{karenprl}.

>From the same formalism as in Sec. \ref{sect:perturbation}, the
free fermion Green's function is now given by \begin{equation}
G_{k}(\sigma)=\left\{ \begin{array}{ll}
\exp[(\beta+\sigma)(\mu-\hbar^{2}k^{2}/2m)], & -\beta<\sigma<0,\\
-\exp[\mu\sigma-\sigma\hbar^{2}k^{2}/2m], & \;\;\;0<\sigma<\beta,\end{array}\right.\end{equation}
so the integral for $\Gamma(k,\sigma,r)$, Eq.~(\ref{Gamma-2}),
gives \begin{equation}
\Gamma(k,\sigma,r)=-ne^{-\sigma(\beta-\sigma)\hbar^{2}k^{2}/2m\beta}e^{-mr^{2}/(2\hbar^{2}\beta)}e^{-ikr\sigma/\beta}.\label{eq:gamma_res}\end{equation}

Substituting Eq.~(\ref{eq:gamma_res}) into Eqs.~(\ref{eq:the_diagram})
and (\ref{delta_g_exchange}) we obtain (see Appendix \ref{append:integrals-Tonks})
\begin{eqnarray}
\Delta g_{d}^{(2)}(r) & = & \frac{2\tau n|r|}{\gamma}e^{-n^{2}\tau r^{2}/2}-\frac{4}{n\gamma}\delta(r),\label{eq:direct_exchange}\\
\Delta g_{e}^{(2)} & = & \frac{4}{n\gamma}\delta(r),\label{eq:exchange}\end{eqnarray}

The only effect of the exchange contribution $\Delta g_{e}^{(2)}$
is to cancel the delta-function in the direct contribution. 
This leaves us with the following result for the pair correlation
function in the regime of high-temperature fermionization ($1\ll\tau\ll\gamma^{2}$):
\begin{equation}
g^{(2)}(r)=1-\left[1-4\sqrt{\frac{\pi\tau}{\gamma^{2}}}\left(\frac{r}{\Lambda_{T}}\right)\right]e^{-(r\sqrt{2\pi}/\Lambda_{T})^{2}}.\label{TGb}\end{equation}
 In the limit $r\rightarrow0$ this leads to perfect antibunching,
$g^{(2)}(0)=0$, while the small finite corrections (as in Ref.~\cite{karenprl},
$g^{(2)}(0)=2\tau/\gamma^{2}$) are reproduced at order $\gamma^{-2}$.
The correlation length associated with the Gaussian decay of correlations
in Eq.~(\ref{TGb}) is given by thermal de Broglie wavelength $\Lambda_{T}=\sqrt{4\pi/(\tau n^{2})}$.
For not very large $\gamma$, the correlations do not decay in a simple
way, but instead show an anomalous, non-monotonic behavior with a
global maximum at at $r_{\max}\simeq\gamma/2\tau n$. This originates
from the effective Pauli-like blocking at short range and thermal
bunching {[}$g^{(2)}(r)>1$] at long range. As $\gamma$ is increased
the position of the maximum diverges and its value approaches 1 in
a non-analytical way $g^{(2)}(r_{\max})\simeq1+(4\tau/\gamma^{2})\exp(-\gamma^{2}/8\tau)$.

\begin{figure}
 \includegraphics*[width=8cm]{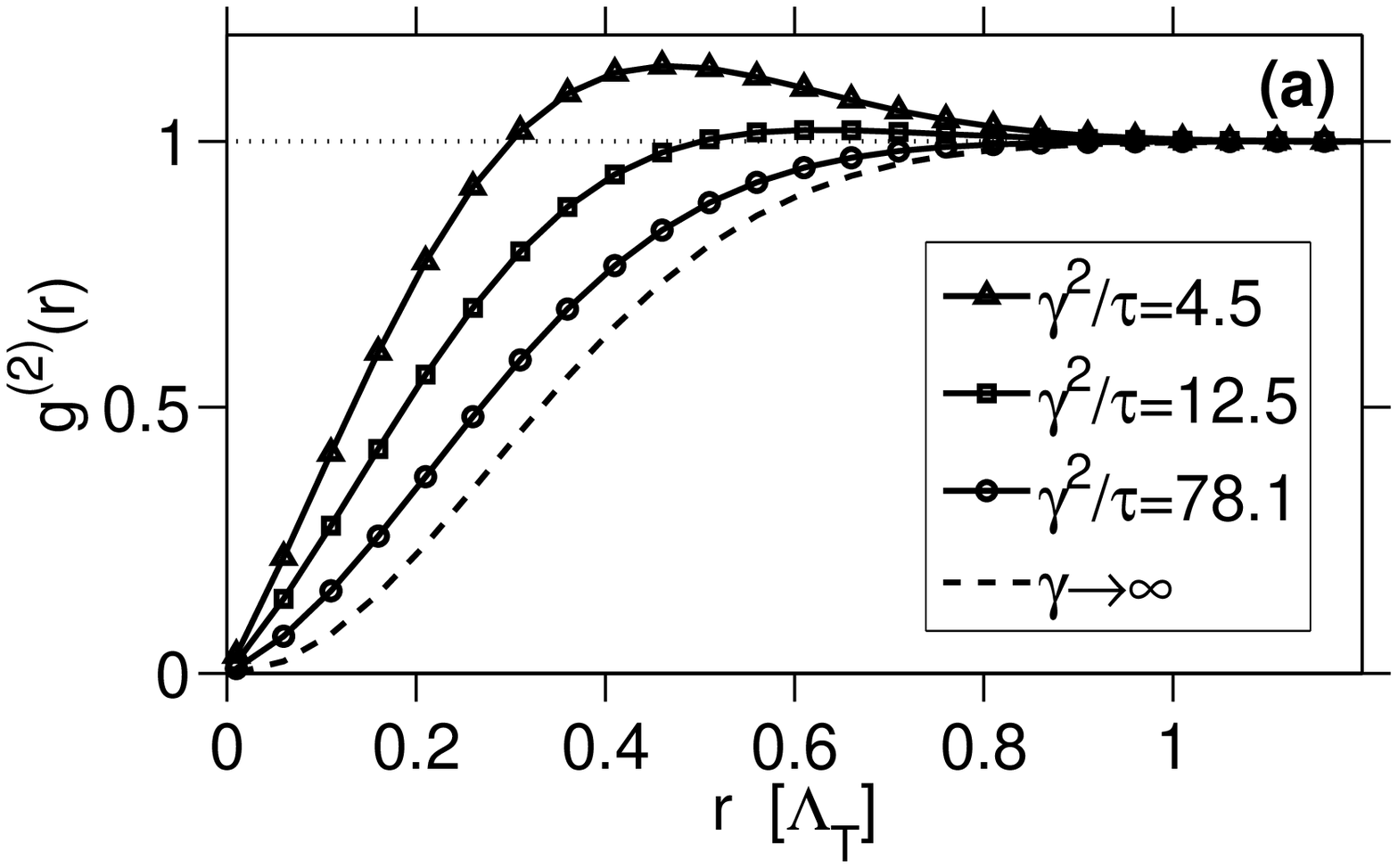}\\
\includegraphics*[width=8cm]{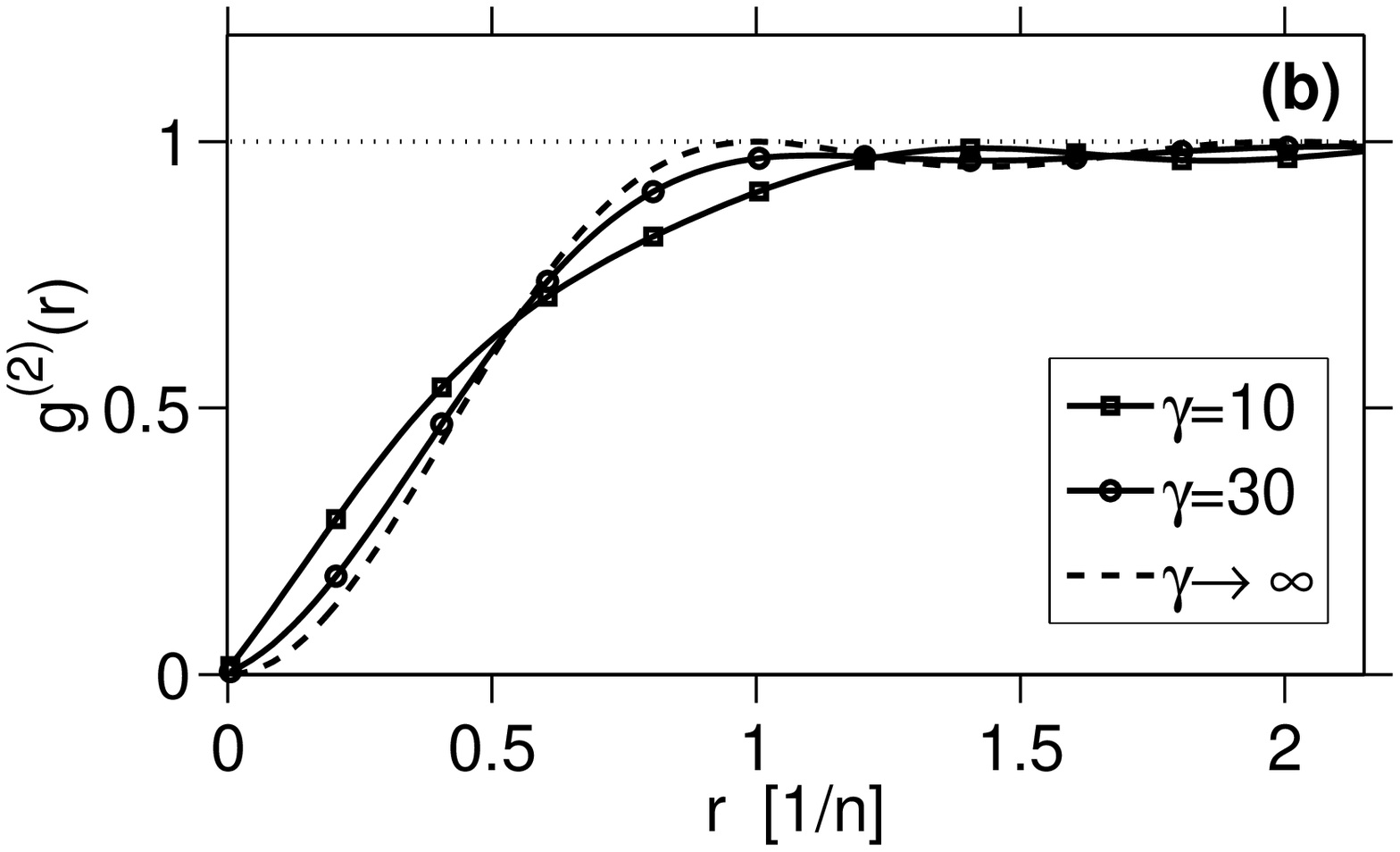}

\caption{Nonlocal pair correlation $g^{(2)}(r)$ as a function of the relative
distance $r$ in the strongly interacting regime, $\gamma\gg1$: (a)
regime of high-temperature {}``fermionization\textquotedblright
, $1\ll\tau\ll\gamma^{2}$, Eq.~(\protect\ref{TGb}), with $r$
in units of the thermal de Broglie wavelength $\Lambda_{T}=\sqrt{4\pi/(\tau n^{2})}$;
(b) low temperature Tonks-Girardeau regime, Eq.~(\protect\ref{TGa}),
for $\tau=0.01$, with $r$ in units of mean interparticle separation
$1/n$}

\label{TGfig}
\end{figure}

Figure~\ref{TGfig}(a) shows a plot of Eq.~(\ref{TGb}) for various
ratios of $\gamma^{2}/\tau$. For a well-pronounced global maximum,
moderate values of $\gamma^{2}/\tau$ are required (such as $\gamma^{2}/\tau\simeq5$,
with $\tau=8$, $\gamma=6$), and these lie near the boundary of validity
($\gamma^{2}/\tau\gg1$) for our perturbative result in the high-temperature
fermionization regime. Exact numerical calculations described in Ref.~\cite{drummond-canonical-gauge},
and in more detail below in Sec.~\ref{sect:HTFxover} do, however,
show qualitatively similar global maxima.

\subsection{Zero- and low-temperature (Tonks-Girardeau) regime}

\label{sect:strong-Tonks}

At $T=0$ the procedure is straightforward~\cite{cherny_brand2}
and yields the known \cite{korepin-book,cherny_brand2} result \begin{gather}
g_{T=0}^{(2)}(r)=1-\frac{\sin^{2}(\zeta)}{\zeta^{2}}-\frac{4}{\gamma}\frac{\sin^{2}(\zeta)}{\zeta^{2}}-\frac{2\pi}{\gamma}\frac{\partial}{\partial\zeta}\frac{\sin^{2}(\zeta)}{\zeta^{2}}\notag\\
+\frac{2}{\gamma}\frac{\partial}{\partial\zeta}\left[\frac{\sin(\zeta)}{\zeta}\int\nolimits _{-1}^{1}dt\sin(\zeta t)\ln\frac{1+t}{1-t}\right],\label{TGa}\end{gather}
where $\zeta\equiv\pi nr$. The last term here diverges logarithmically
with $\zeta$ and can be regarded as a first order perturbation correction
to the fermionic inverse square power law. Accordingly, Eq.~(\ref{TGa})
is valid for $\zeta\ll\exp(\gamma)$.

At temperatures well below quantum degeneracy, $\tau\ll1$, finite
temperature corrections to Eq.~(\ref{TGa}) are obtained using a
Sommerfeld expansion around the zero temperature Fermi-Dirac distribution
for the quasi-momenta. For $rn\ll\tau^{-1}$ this gives an additional
contribution of $\tau^{2}\sin^{2}(\pi nr)/12\pi^{2}$ to the right
hand side of Eq.~(\ref{TGa}), which is negligible compared to the
$T=0$ result as $\tau\ll1$. At $r=0$, Eq.~(\ref{TGa}) gives perfect
antibunching $g^{(2)}(0)=0$, which corresponds to a fully {}``fermionized\textquotedblright\ 1D
Bose gas, where the strong inter-atomic repulsion mimics the Pauli
exclusion principle for intrinsic fermions. By extending the perturbation
theory to include terms of order $\gamma^{-2}$ we can reproduce the
known result for the local pair correlation at zero temperature $g^{(2)}(0)=4\pi^{2}/3\gamma^{2}$
\cite{karenprl,gangardt-correlations}.

In Fig.~\ref{TGfig}(b) we plot the function $g^{(2)}(r)$, Eq.~(\ref{TGa}),
for various $\gamma$. According to the physical interpretation of
the pair correlation function $g^{(2)}(r)$, its oscillatory structure,
and hence the existence of local maxima and minima at certain finite
values of $r$, implies that there exist more and less likely separations
between the pairs of particles in the gas. This can be interpreted
as a quasi-crystalline order (with a period of $\sim1/n$) in the
two-particle sector of the many-body wave function even though the
density of the gas is uniform.

The oscillatory behavior of the pair correlation in this strongly
interacting regime is similar to Friedel oscillations in the density
profile of a 1D interacting electron gas with an impurity \cite{Friedel}.
We also mention that our derivation of Eq.~(\ref{TGa}) is equally
valid for strong attractive interactions, i.e., when $\gamma<0$ and
$|\gamma|\gg1$, and therefore it describes the pair correlations
in a metastable state known as super-Tonks gas \cite{super-Tonks}.

\subsection{Numerical results}

\label{subsect:HTF-numerix}

\begin{figure}
 \includegraphics*[width=8cm]{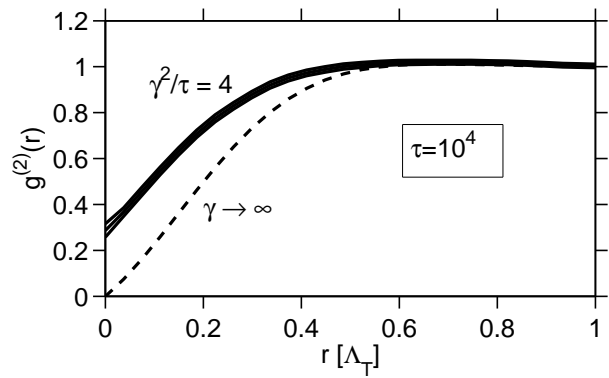}

\caption{Behavior on the verge of the high-$T$ fermionization regime for $\gamma^{2}/\tau=4$.
The dashed line is Eq. (\protect\ref{TGb}).}

\label{HTF_verge}
\end{figure}

Numerical calculations with the gauge-$P$ method are able to reach
only the low-$\gamma$ (or, equivalently, high $\tau$) edge of the
high-temperature fermionization regime, however a comparison with
Eq. (\ref{TGb}) is instructive. In Fig.~\ref{HTF_verge} we see
that the length scale on which antibunching occurs is still qualitatively
given by Eq. (\ref{TGb}) while any discrepancies are of the same
size as at $r=0$. This is actually a general feature in all the parameter
regimes explored by the numerics. Overall, the discrepancy between
the $1/\gamma$ perturbation expansions (\ref{DC}), (\ref{DQ}),
(\ref{TGb}), and the exact behavior of $g^{(2)}(r)$ at nonzero $r$
is roughly the same as at $r=0$. Since a calculation of $g^{(2)}(0)$
\cite{karenprl} from the exact solution of the Yang-Yang integral
equations \cite{yangyang1} is usually more straightforward to evaluate
than the full stochastic calculation of $g^{(2)}(r)$, it can serve
as a useful guide to whether a numerical calculation is warranted
or not.

\section{Classical/Fermionization transition and correlation maxima}

\label{sect:HTFxover}

\begin{figure}
 \includegraphics*[width=8cm]{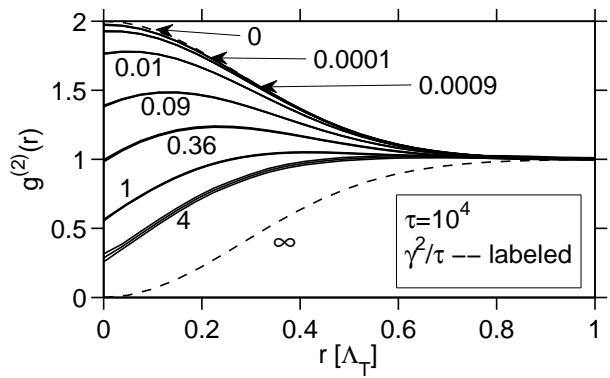}

\caption{Crossover from decoherent classical to high temperature fermionization
regimes at high temperature.}

\label{HTFDC}
\end{figure}

Figure~\ref{HTFDC} shows the behavior in the transition region between
the decoherent classical and high temperature fermionization regimes
(found with the gauge-$P$ numerical method), when one is far above
the degeneracy temperature $T_{d}$. One sees the appearance of a
maximum in the correlations at finite range as the transition is approached.
As pointed out in Sec.~\ref{sect:fermionization}, this arises from
an interplay of thermal bunching and repulsive antibunching on comparable
scales. A comparison of relevant length scales indicates that the
$\tau\approx\gamma^{2}$ here corresponds to $\Lambda_{T}\sim a_{1D}$,
where $a_{1D}$ is the {}``1D scattering length'' that describes
the asymptotic behavior of the wave function in two-body scattering.

\begin{figure}
 \includegraphics*[width=8cm]{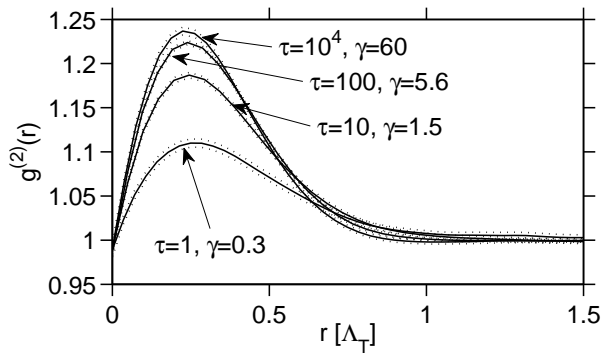}

\caption{Situation when $T>T_{d}$ and the local second-order coherence is
apparently unity. All curves plotted correspond to parameter values
for which $g^{(2)}(0)=1$ in the crossover region between the classical
decoherent and high-temperature fermionized gas. The dots (rather
than triple lines here, for clarity) indicate $1\sigma$ error bars.}

\label{g2is1}
\end{figure}

An interesting behavior occurs in the crossover regime when $\gamma^{2}/\tau\simeq0.1-0.4$.
Here we can have $g^{(2)}(0)=1$ just like in the quasi-condensate
or {}``Gross-Pitaevskii\textquotedblright\ regime, indicating local
second-order coherence. However, unlike the quasi-condensate regime,
the non-local correlations on length scales of $\sim\Lambda_{T}$
are \emph{not} coherent, and in fact appreciably bunched. This is
shown in Fig.~\ref{g2is1}. It is a symptom of the broader correlation
maximum phenomenon.

\begin{figure}
 \includegraphics*[width=8cm]{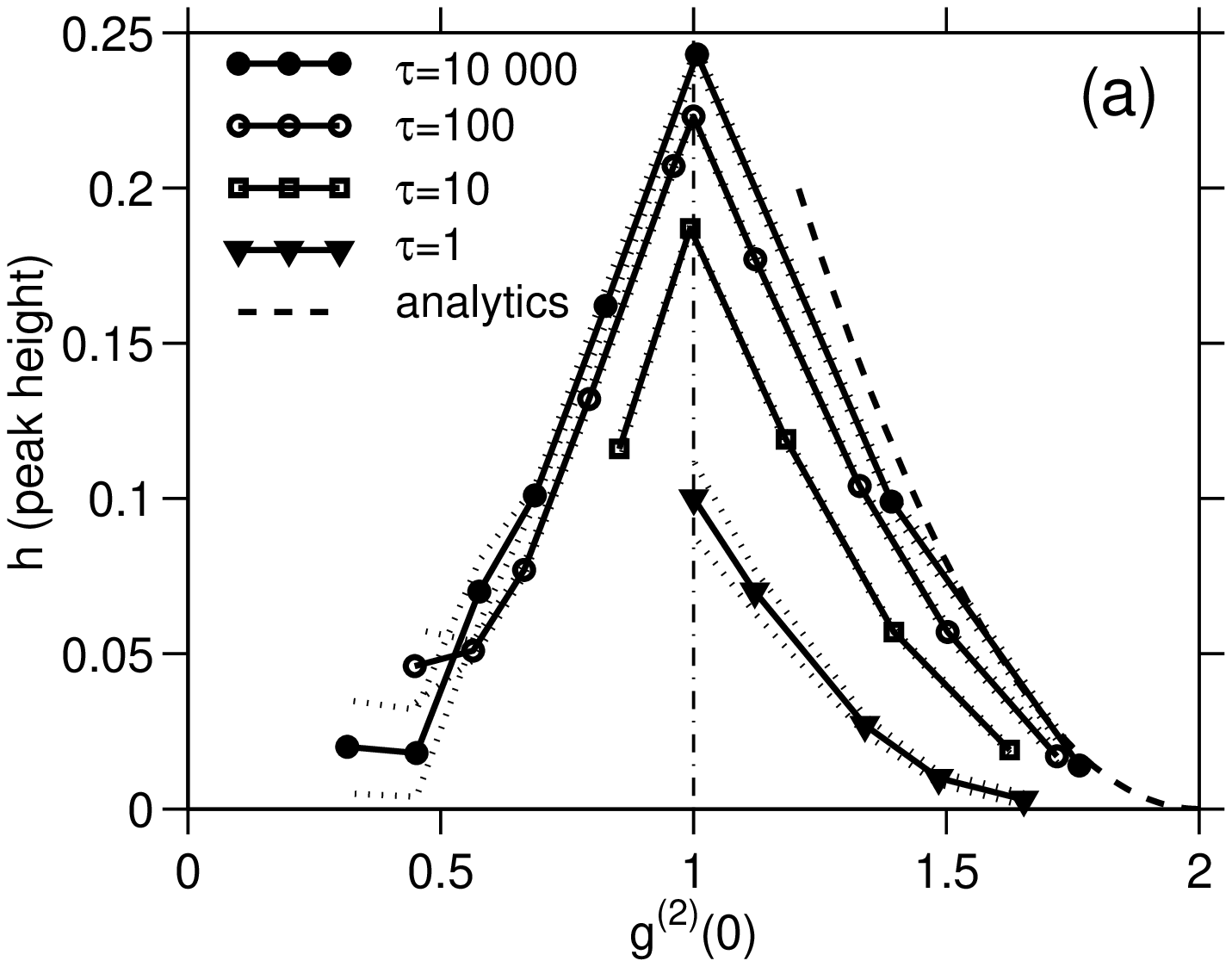}\\
\includegraphics*[width=8cm]{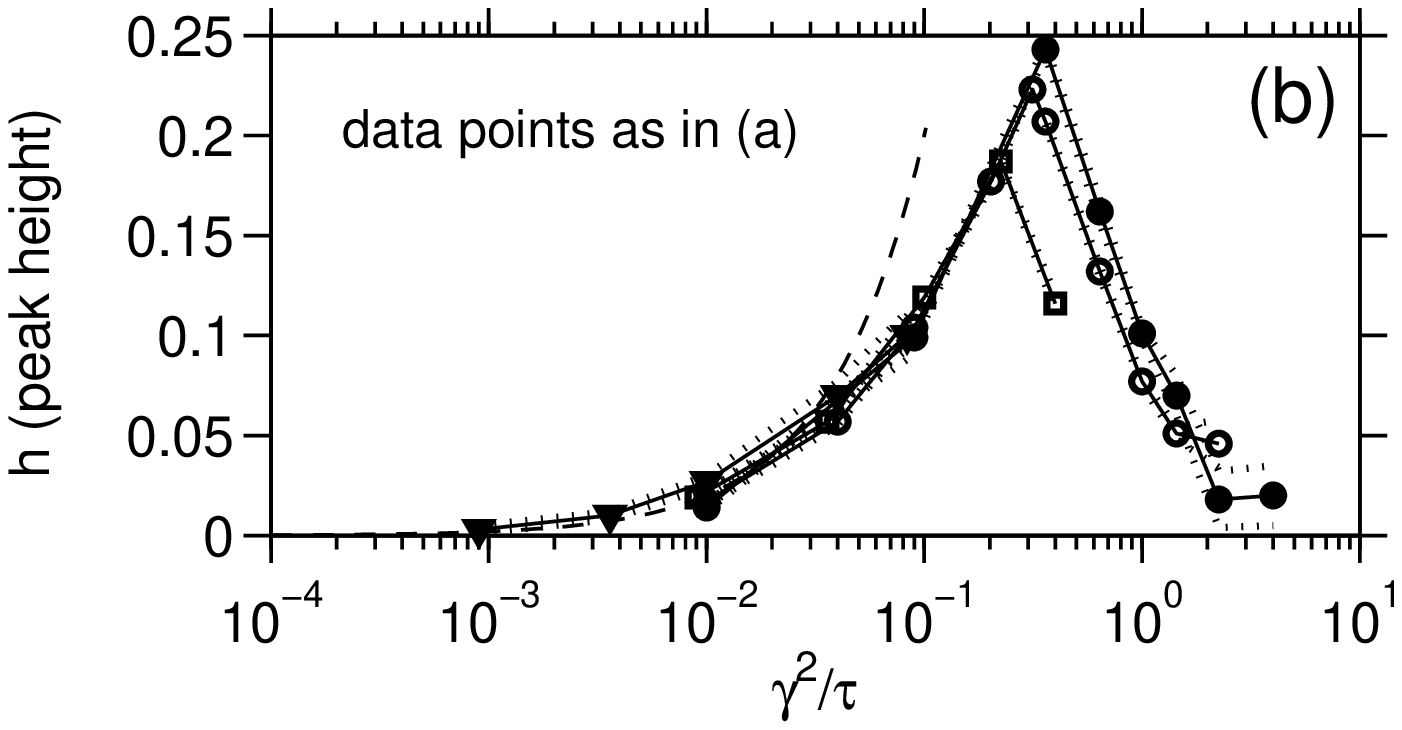}

\caption{Heights of the anomalous peak of $g^{(2)}(r)$ that occurs at nonzero
$r_{\max}$, for different values of $\tau$, as functions of $g^{(2)}(0)$
-- (a) and $\gamma^{2}/\tau$ -- (b). The height is taken to be $h\equiv g^{(2)}(r_{\max})-g^{(2)}(0)$
at high temperatures when $g^{(2)}(0)>1$, and $h\equiv g^{(2)}(r_{\max})-1$
when $g^{(2)}(0)<1$. The two regimes are separated by the dot-dashed
vertical line in (a). Analytic results from Eq.~(\protect\ref{DC})
in the decoherent quantum regime are shown as a dashed line. Dots
(rather than triple lines here, for clarity) indicate $1\sigma$ error
bars on the numerical results.}

\label{peakheights}
\end{figure}

The height of this maximum for more general parameters is shown in
Fig.~\ref{peakheights} as a function of both $g^{(2)}(0)$ and $\gamma^{2}/\tau$.
One sees that this behavior is well pronounced in the crossover between
high temperature fermionization and decoherent classical regimes,
peaking when $g^{(2)}(0)\simeq1$ (a situation shown also in Fig.~\ref{g2is1}),
or, equivalently, $\gamma^{2}\sim0.3\tau$. As one reaches degenerate
temperatures, the maximum peak height is reduced, and presumably disappears
completely by the time the quasi-condensate regime is reached by going
to smaller values of $\gamma$. Although we were unable to numerically
reach the relevant quasi-condensate region for $\tau<1$, a more refined
numerical setup that improves the importance sampling or the $\mu(T)$
trajectory described in Appendix~\ref{append:numerix} may allow
this.

\section{Numerical limitations}

\label{sect:numerical-limitations}

\begin{figure}
 \includegraphics*[width=8cm]{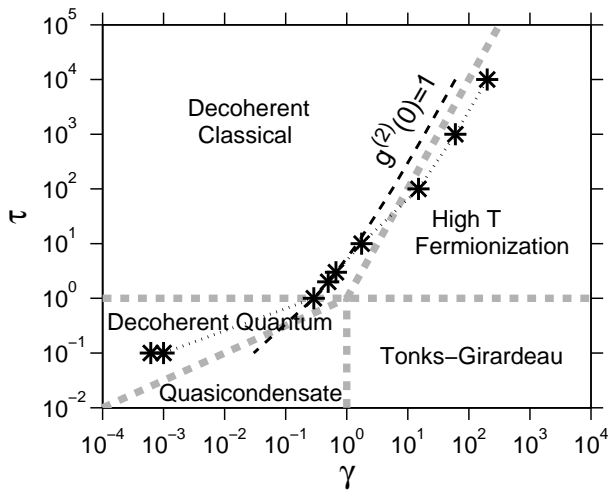}

\caption{Regimes and their numerical accessibility: the asterisks indicate
the lowest $\tau$ and highest $\gamma$ reachable using the gauge-$P$
method as described in Appendix~\protect\ref{append:numerix}. The
dark dashed line indicates the point at which $g^{(2)}(0)=1$. }

\label{Fig:numerixlimit}
\end{figure}

Figure~\ref{Fig:numerixlimit} shows the regime that was accessible
using the relatively straightforward numerical scheme that was employed
here, and detailed in Appendix~\ref{append:numerix}. (It is the
region above and to the left of the asterisks). In particular, one
sees that of the physical regimes described in previous sections,
the decoherent classical, as well as parts of the decoherent quantum
and high-temperature fermionization regimes were accessible, while
the quasi-condensate and Tonks-Girardeau regimes were not.

The principal difficulty that is encountered, generally speaking,
is the growth of statistical noise with increasing $\beta$, i.e.
decreasing $\tau$, which eventually prevents one from obtaining values
of $g^{(2)}(r)$ with a useful resolution. This arises in two different
ways depending on the region of interest.

Firstly, in the strongly interacting (fermionized) region, one needs
a correspondingly large coupling constant $g\propto\gamma$ which
leads to a relative increase of the importance of the noise terms of
the $d\alpha_{j}^{(\nu)}/d\beta$ equations in (\ref{Gequations}).
This leads to large statistical uncertainty in the
$\alpha_{j}^{(\nu)}$ themselves or to the weight $\Omega$ whose
evolution depends on them. The upshot is that the inverse
temperature $\beta$ at which the noise becomes unmanageable becomes
smaller and smaller as $\gamma$ grows. Technical improvements are
unlikely to make a large dent in the problem in the fermionized
regime because it ultimately stems from the fact that coherent
states are no longer a good basis over which to expand the density
matrix. They are not close to the preferred eigenstates of the
system. Instead, one can think of constructing a phase-space
distribution that uses a non-coherent-state basis, for example, a
Gaussian basis \cite{CorneyDrummond}. This general approach -
together with symmetry projections - has been utilized in
successfully calculating ground state properties of the strongly
correlated fermionic Hubbard model \cite{AimiImada}.

Secondly, in the low $\gamma$ and $\tau$ region, one has a different
underlying source of statistical uncertainty. The longest relevant
length here is either the coherence length $l_{\phi}$ or the healing
length $\xi$, and for correct calculations in the large uniform gas
one must simulate a system of a total size appreciably greater than
these lengths. This in turn imposes a minimal total particle number
\begin{equation}
N\gtrsim\mathrm{max}\left[\mathcal{O}(4/\tau),\mathcal{O}(2/\sqrt{\gamma})\right].\label{Nminimum}\end{equation}
The thermal initial conditions of Eq. (\ref{G0}) lead to variation
in $N$ among trajectories, and since the Gibbs factor $K$ (see Eq.
(\ref{GibbsK})\,) grows linearly or faster with $N$, one also obtains
a growing variation of $K(\vec{v})$. This enters the $d\Omega$ of
Eq. (\ref{Gequations}) and leads to a spread of the weights $\Omega(t)$
that grows rapidly (note the exponential growth of $\Omega$) with
increasing $N$. However because of the long length scales, via (\ref{Nminimum}),
large $N$ is needed to make accurate calculations when $\tau$ or
$\gamma$ are much smaller than one. The end result is domination
of the whole calculation by one or a few trajectories with the highest
weight, for all realistic ensemble sizes $\mathcal{S}$.

As a corollary, significantly lower temperatures, even down to the
quasi-condensate regime, \emph{are} accessible at small $\gamma$ if
one is prepared to sacrifice the assumption of an infinite-sized gas
and consider periodic boundary conditions on some length $L$ that is
smaller than or comparable to the coherence/healing lengths. This
approach was taken, e.g., in \cite{Carusotto}. This stops the rise
of overall particle number, hence one has a much smaller spread of
Gibbs factors $\Omega$ among the trajectories, and in the final
analysis -- reduced statistical uncertainty. Such calculations are
no longer as general, though, and are not considered in this paper.

We would like to point out that the limitation in this regime may
be overcome or alleviated if the rather simplistic importance sampling
used in the numerical method were to be improved. The leading candidate
is an improved importance sampling algorithm, possibly using a Metropolis
sampling procedure, as outlined at the end of Appendix~\ref{append:preweight}.

Finally, it is also possible that a more refined choice of $\mu(\beta)$
(considered in Appendix~\ref{append:mu}) may lead to somewhat improved
coverage of the parameter space in general.

\section{Overview and conclusion}

\label{sect:conclusions}


In conclusion, we have surveyed the behavior of the spatial two-particle
correlation function in a repulsive uniform 1D Bose gas. We have analyzed
numerically the pair correlation functions for all relevant length
scales, with the exception of several low-temperature transition regions
(see Fig.~\ref{Fig:numerixlimit} below the asterisks) which were
not accessible by the numerical scheme we employed. Approximate analytic
results and methods have been presented for parameters deep within
all the major physical regimes. The key features of this behavior
include:
\begin{itemize}
\item Thermal bunching with $g^{(2)}(0)\simeq2$ and Gaussian drop-off at
ranges $\Lambda_{T}$ in the classical decoherent regime.
\item Exponential drop-off of correlations from $g^{(2)}(0)\simeq2$ at
ranges $l_{\phi}$ in the decoherent quantum regime, along with Gaussian-like
rounding at shorter ranges $\sim\Lambda_{T}$.
\item Suppressed density fluctuations with $g^{(2)}(0)\simeq1$ and exponential
decay at ranges of the healing length $\xi$ in the quasi-condensate
regime.
\item Antibunching with $g^{(2)}(0)<1$ and Gaussian decay at ranges $\Lambda_{T}$
in the high-temperature fermionization regime.
\item Antibunching with $g^{(2)}(0)<1$ and oscillatory decay on ranges
of the mean interparticle separation $1/n$ in the Tonks-Girardeau
regime.
\item Bunching at a range of $\sim0.3\Lambda_{T}$ in the crossover between
classical and fermionized regimes around $\gamma^{2}\sim0.3\tau$.
\end{itemize}

Let us consider the regimes in turn, starting from the classical decoherent
gas, then going anti-clockwise in Fig.~\ref{Fig:numerixlimit}. The
classical decoherent gas is well approximated by Boltzmann statistics
and is dominated by thermal fluctuations. The pair correlation function
shows typical thermal bunching and a Gaussian decay, with the correlation
length given by the thermal de Broglie wavelength $\Lambda_{T}$.

As one reduces the temperature, the gas becomes degenerate, the thermal
de Broglie wavelength becomes larger than the mean interparticle separation
and loses its relevance. The correlation length increases and one
enters into the decoherent quantum regime. Here, the dominant behavior
of the gas is the ideal Bose gas bunching, $g^{(2)}(0)\simeq2$, with
large density fluctuations that decay exponentially on the length
scale given by the phase coherence length $l_{\phi}$. Notably, the
exponential behaviour starts to appear well above degeneracy first
in the long-distance tails, being visible even around $\tau\sim10$
as in Fig.~\ref{lowgam}.

Reducing the temperature even further, while still at $\gamma\ll1$,
one enters into the quasi-condensate regime, in which the density
fluctuations become suppressed and $g^{(2)}(0)\simeq1$. In the
hotter sub-regime dominated by thermal fluctuations, the pair
correlation shows weak bunching, $g^{(2)}(0)>1$, while in the colder
sub-regime dominated by quantum fluctuations one has weak
antibunching, $g^{(2)}(0)<1$. In both cases the pair correlation
decays on the length scale of the healing length $\xi$.

We now move to the right on Fig.~\ref{Fig:numerixlimit}, into the
regime of strong interactions, while staying at temperatures well
below quantum degeneracy, $\tau\ll1$. This is the Tonks-Girardeau
regime, in which the density fluctuations get further suppressed due
to strong interparticle repulsion. Antibunching increases and one
approaches $g^{(2)}(0)=0$ due to fermionization. The only relevant
length scale here is the mean interparticle separation, $1/n$, and
the pair correlation function decays on this length scale with some
oscillations.

We next move up on Fig.~\ref{Fig:numerixlimit}, to higher temperatures,
and enter the regime of high-temperature fermionization. At short
range, the pair correlation here is still antibunched due to strong
interparticle repulsion, however, thermal effects start to show up
on the length scale of $\Lambda_{T}$. As a result of these competing
effects, the nonlocal pair correlation develops an anomalous peak,
corresponding to bunching at-a-distance, with $g^{(2)}(r_{\max})>1$,
beginning around $\tau\sim\gamma^{2}/2$.

As we increase the temperature even further, the thermal effects start
to dominate over interactions and the antibunching dip gradually disappears.
At temperatures $\tau\sim\gamma^{2}$ we observe a crossover back
to the classical decoherent regime.

Our results provide new insights into the fundamental understanding
of the 1D Bose gas model through many-body correlations. Calculation
of these non-local correlations is not accessible yet through the
exact Bethe ansatz solutions. We expect that our theoretical predictions
will serve as guidelines for future experiments aimed at the measurement
of nonlocal pair correlations in quasi-1D Bose gases.



\begin{acknowledgments}
AGS, MJD, PDD and KVK acknowledge fruitful
discussions with A. Yu. Cherny and J. Brand, and the support of this
work by the Australian Research Council. DMG acknowledges support
by EPSRC Advanced Fellowship EP/D072514/1. PD was supported by the
European Community under the contract MEIF-CT-2006-041390. KVK, PD
and DMG thank IFRAF and the Institut Henri Poincare--Centre Emile
Borel for support during the 2007 Quantum Gases workshop in Paris
where part of this work was completed. LPTMS is a mixed research unit No. 8626 of CNRS and Universit\'{e} Paris-Sud.
\end{acknowledgments}

\appendix


\section{Technical appendix for the gauge-$P$ calculations}

\label{append:numerix}

\subsection{Instability of the stochastic equations and its removal with a stochastic
gauge}

\label{append:gauge}

\begin{figure}
 \includegraphics*[width=8cm]{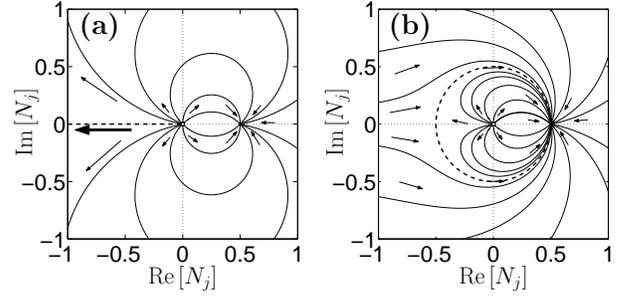}

\caption{Deterministic phase space for Stratonovich for of the $dN_{j}$ equation,
when $\mu_{e}=0$. \textbf{(a)}: ungauged, \textbf{(b)}: using the
gauge (\protect\ref{mingauge}). The moving singularity in \textbf{(a)}
is shown with a large arrow, the attractor in \textbf{(b)} at $|N_{j}|=N_{a}$
with a thick dashed line.}

\label{FigGauge}
\end{figure}

A straightforward application of the ungauged diffusion Eqs. (\ref{ppequations})
is foiled by the presence of an instability in the $d\alpha_{j}^{(\nu)}/d\beta$
equations. We can see this if we first consider the evolution of $N_{j}$
and discard the noise and kinetic-energy parts of the equation. Taking
the deterministic part from the Stratonovich calculus which is used
for our numerics (this introduces the $1/2$ term below), one has
\begin{equation}
\frac{\partial N_{j}}{\partial\beta}\sim N_{j}\left[\mu_{e}-\frac{g}{\Delta x}\left(N_{j}-\frac{1}{2}\right)\right].\label{deterministic}\end{equation}
There are stationary points at the vacuum $N_{j}=0$ and at $N_{j}=N_{a}=1/2+\mu_{e}\Delta x/g$,
with the more positive stationary point (usually $N_{a}$) being an
attractor, and the more negative a repellor {[}see Fig.~\ref{FigGauge}
(a)\,]. The deterministic evolution is easily solved, and starting
from a time $\beta_{0}$ gives later evolution as \begin{equation}
N_{j}(\beta)=\frac{N_{a}N_{j}(\beta_{0})}{N_{j}(\beta_{0})+(N_{a}-N_{j}(\beta_{0}))e^{-\mu_{e}(\beta-\beta_{0})}}.\end{equation}
If has a negative $N_{j}(\beta_{0})$, which is possible due to the
action of the noises $\zeta$, then at a later time \begin{equation}
\beta_{\text{sing}}=\beta_{0}+\frac{1}{\mu_{e}}\ln\left(1-\frac{N_{a}}{N_{j}(\beta_{0})}\right),\end{equation}
the solution has diverged to negative infinity. This behavior of the
deterministic part of the equations is known as a {}``moving singularity\textquotedblright\ and
is a well-known indicator of non-vanishing boundary terms when an
integration-by-parts is performed on the operator equation (\ref{differential})
\cite{Gilchrist,deuar-thesis}. It implies that the FPE (\ref{ppfpe})
is not fully equivalent to quantum mechanics.

The use of a stochastic gauge to remove this kind of instability has
been described in \cite{deuar-drummond-2006}, and in more detail
in \cite{deuar-thesis}. The gauge identity, Eq. (\ref{gaugeidentity}),
can be used on Eq. (\ref{differential}) to introduce an arbitrary
modification to the deterministic evolution (arising from first order
derivative terms) for the price of additional diffusion in the weight
$\Omega$. Since the gauge identity is zero, we can add an arbitrary
multiple of it to Eq. (\ref{differential}). In particular, if we
add \begin{eqnarray}
0 & = & \int G(\vec{v})\sum_{j}\left\{ \frac{\mathcal{G}_{j}^{2}\Omega^{2}}{2}\frac{\partial^{2}}{\partial\Omega^{2}}\right.\\
 &  & \hspace*{-2em}\left.+i\mathcal{G}_{j}\sqrt{\frac{g}{2\Delta x}}\,\sum_{\nu}\alpha_{j}^{(\nu)}\frac{\partial}{\partial\alpha_{j}^{(\nu)}}\left(\Omega\frac{\partial}{\partial\Omega}-1\right)\right\} \widehat{\Lambda}\, d^{4M+2}\notag\end{eqnarray}
with arbitrary functions $\mathcal{G}_{j}(\vec{v},\beta)$, and perform
the subsequent steps as before, then the diffusion matrix in the resulting
FPE remains positive semidefinite (no negative eigenvalues), and the
resulting Ito diffusion equations of the samples become \begin{eqnarray}
\frac{d\alpha_{j}^{(\nu)}}{d\beta} & = & \frac{1}{2}\left(\mu_{e}+\frac{\hbar^{2}\nabla^{2}}{2m}-\frac{gN_{j}}{\Delta x}\right)\alpha_{j}^{(\nu)}\notag\\
 &  & +i\alpha_{j}^{(\nu)}\left[\zeta_{j}^{(\nu)}(\beta)-\mathcal{G}_{j}\right]\sqrt{\frac{g}{2\Delta x}}\,,\label{Gexplicitequations}\\
\frac{d\Omega}{d\beta} & = & \Omega\left[-K(\vec{v})+\sum_{j}\mathcal{G}_{j}\sum_{\nu}\zeta_{j}^{(\nu)}(\beta)\right],\notag\end{eqnarray}
instead of (\ref{ppequations}). The $\alpha_{j}$ equations are modified
and compensating correlated noises have been added to the $\Omega$
equation.

We now wish to choose the functions $\mathcal{G}_{j}$, called stochastic
gauges, so that the instability is removed, keeping also in mind the
goal of keeping the (now unbiased) statistical uncertainty manageable.
Heuristic guidelines for choosing gauges have been investigated in
detail in \cite{deuar-thesis}. Several choices for a single-mode
system were also investigated there in Sec. $9.2$ in terms of resulting
statistical uncertainties. The aim is to remove the real part of $N_{j}$
from the $\alpha_{j}$ equation when it is negative, so as to neutralize
the moving singularity. While for a single mode the {}``radial\textquotedblright\ gauge
was found to give the best performance, later tests that we have performed
on the full multimode ($M\gg1$) 1D gas show that the {}``minimal\textquotedblright\ drift
gauge \begin{equation}
\mathcal{G}_{j}=i\left(\text{Re}N_{j}-|N_{j}|\right)\sqrt{\frac{g}{2\Delta x}}\label{mingauge}\end{equation}
gives better performance for this system. This is because it introduces
the smallest modifications needed to remove the moving singularity,
and hence the smallest noise contributions to the weight $\Omega$.
The weight becomes much more important for multimode systems because
each of the $M$ modes adds its own contribution to it, the total
of which can become large. The phase-space modification for a single
mode for the ungauged Eq. (\ref{deterministic}) and gauged equations
is shown in Fig.~\ref{FigGauge}. One sees that in the {}``classical''
$\mathrm{Re}[N_{j}]\gg\mathrm{Im}[N_{j}]$ region the trajectories
are practically unchanged. The final Ito equations to be integrated
are (\ref{Gequations}). Comparisons to known exact results such as
energy and density \cite{yangyang1}, and $g^{(2)}(0)$ \cite{karenprl}
indicate no deviations beyond what is predicted by the unbiased statistical
uncertainties, Eq. (\ref{uncertainty}), with the new gauged equations.
Such a comparison can be seen in Fig. 2 of Ref. \cite{drummond-canonical-gauge}.

\subsection{Integration procedure}

\label{append:integration} The actual integration is performed using
a split-step semi-implicit method described in \cite{Ian}, which
requires the use of the Stratonovich stochastic calculus. There, it
was shown to be highly superior to other low-order methods in terms
of stability. Although a low order Newton-like method, with the right
choice of variables its performance is remarkably good. High-order
methods such as Runge-Kutta or others suffer from serious complications
when noise is present. In particular, one has to be very meticulous
in tracking down and compensating for all the non-zero correlations
within a single time-step --- these are much more complicated than
the simplest correction terms appearing in the Stratonovich semi-implicit
method used here.

Due to the multiplicative form of the equations (\ref{Gequations}),
it is highly advantageous to use logarithmic variables, which is made
possible if one uses a split-step method. Here, a $\Delta\beta$ timestep
consists of the following four stages: First the interaction part
(containing $g$) is integrated in real space over a time-step $\Delta\beta$.
Second, the fields are Fourier-transformed to $k$-space, giving $\widetilde{\alpha}^{(\nu)}(k)$.
Thirdly the kinetic-energy contributions are integrated over $\Delta\beta$,
and finally one Fourier-transforms back into real space, ready to
start the next timestep. The Stratonovich gauged evolution equations
for the real space stage are \begin{subequations} \label{logequations}
\begin{eqnarray}
\frac{d\ln\alpha_{j}^{(\nu)}}{d\beta} & = & -\frac{g}{2\Delta x}\left(|N_{j}|+i\,\text{Im}N_{j}-\frac{1}{2}\right)\notag\\
 &  & +i\zeta_{j}^{(\nu)}(\beta)\sqrt{\frac{g}{2\Delta x}},\notag\\
\frac{d\ln\Omega}{d\beta} & = & i\sqrt{\frac{g}{2\Delta x}}\sum_{j,\nu}\left(\text{Re}N_{j}-|N_{j}|\right)\zeta_{j}^{(\nu)}(\beta)\label{trueequationsx}\\
 &  & +\frac{g}{2\Delta x}\sum_{j}\left\{ (\text{Re}N_{j}-|N_{j}|)^{2}-N_{j}^{2}+i\text{Im}N_{j}\right\} ,\notag\end{eqnarray}
while for the $k$-space stage they are\begin{eqnarray}
\frac{d\ln\widetilde{\alpha}^{(\nu)}(k)}{d\beta} & = & \frac{1}{2}\left[\mu_{e}-\frac{\hbar^{2}k^{2}}{2m}\right],\label{trueequationsk}\\
\frac{d\ln\Omega}{d\beta} & = & \sum_{k}\left(\mu_{e}-\frac{\hbar^{2}k^{2}}{2m}\right)\widetilde{\alpha}^{+}(k)\widetilde{\alpha}(k).\notag\end{eqnarray}

\subsection{Importance sampling}

\label{append:preweight}

The simulated equations (\ref{Gequations}) include evolution of both
the amplitudes $\alpha_{j}^{(\nu)}$ and weight $\Omega$. This combination
can cause sampling problems for observable estimations, Eq. (\ref{obs}),
when maximum weights occur for very rare trajectories. As it turns
out, this was a serious issue for the majority of calculations reported
here because while the initial distribution (\ref{G0}) samples the
$\beta=0$ system well, this is not necessarily the case during the
later evolution into $\beta\gg0$ that is of most interest. Fortunately,
fairly rudimentary importance sampling was able to deal with this
for a wide range of parameters.

The essence of this approach is to pre-weight trajectories in such
a way that the part of the distribution with maximum weight $\Omega$
coincides with the majority of samples at the target time of interest
$\beta_{t}$, rather than at $\beta=0$. The price paid is that the
$\beta=0$ distribution is then poorly sampled, but this is not important
to us as we are interested rather in the target $\beta_{t}$.

Pre-weighting is made possible because in all observable
calculations (\ref{obs}), the combination $[G(\vec{v})\Omega]$
occurs as a universal common factor in the $\int d\vec{v}$ integral.
Hence, if we manually scale the weight $\Omega$ by some factor
$F(\vec{v})$ of our choice:
$\Omega\rightarrow\Omega^{\prime}F(\vec{v})$, and simultaneously
rescale the distribution according to $G(\vec{v})\rightarrow
G^{\prime}(\vec{v})/F(\vec{v})$, then with $\Omega^{\prime}$ and
$G^{\prime}$ one obtains exactly the same results in the
infinite-number-of-samples limit as with $G\Omega$. However, the
actual samples are differently distributed, which is advantageous
for finite sample numbers. To reduce the weight sampling problem,
one wants to make such a modification $F(\vec{v})$ that both
$G^{\prime}(\vec{v})$ and $\Omega^{\prime}G^{\prime}(\vec{v})$ peak
in the same region of the phase space of ${\vec{v}}$.

To proceed, it is convenient to consider Fourier-transformed variables
in $k$-space, where the non-interacting evolution can be easily exactly
solved. Define then \end{subequations} \begin{equation}
\widetilde{\alpha}_{k}^{(\nu)}=\frac{1}{\sqrt{M}}\sum_{j}e^{-ikx_{j}}\alpha_{j}^{(\nu)}=\left\{ \begin{array}{cl}
\widetilde{\alpha}_{k}, & \text{ if }\nu=1,\\
\widetilde{\alpha}_{k}^{+}, & \text{ if }\nu=2,\end{array}\right.\end{equation}
where $k$ takes on discrete values from $-\pi/\Delta x$ to $\pi/\Delta x$.
The {}``naive\textquotedblright\ initial distribution (\ref{G0})
then becomes \begin{equation}
G_{0}(\vec{v})=\delta^{2}(\ln\Omega)\prod_{k}\delta^{2}(\widetilde{\alpha}_{k}-(\widetilde{\alpha}_{k}^{+})^{\ast}\,)\frac{e^{-|\widetilde{\alpha}_{k}|^{2}/\overline{n}_{x}}}{\pi\,\overline{n}_{x}}.\end{equation}
This is a thermal distribution which is uniform over all $k$. The
ideal gas (i.e. $g=0$) evolution of equations (\ref{Gequations})
then leads to \begin{eqnarray}
\widetilde{\alpha}_{k}^{(\nu)}(\beta) & = & \widetilde{\alpha}_{k}(0)\exp\left[\left(\mu(\beta)-\frac{\hbar^{2}k^{2}}{2m}\right)\frac{\beta}{2}\right],\label{idealgassol}\\
\ln\Omega(\beta) & = & \sum_{k}\left(|\widetilde{\alpha}_{k}(\beta)|^{2}-|\widetilde{\alpha}_{k}(0)|^{2}\right),\notag\end{eqnarray}
where \begin{equation}
\widetilde{\alpha}_{k}(0)=\sqrt{\overline{n}_{x}}\,\eta_{k},\end{equation}
with $\eta_{k}$ being independent complex Gaussian noises with variance
unity, $\langle\eta_{k}^{\ast}\eta_{k^{\prime}}\rangle_{\mathcal{S}}=\delta_{kk^{\prime}}$.
One can see that (\ref{idealgassol}) is not necessarily anywhere
near a well-sampled ideal gas Bose-Einstein distribution at temperature
$\beta$, which would have \begin{eqnarray}
\widetilde{\alpha}_{k}^{(\nu)}(\beta) & = & \sqrt{n_{k}^{\mathrm{id}}(\beta)}\ {\eta_{k},}\notag\\
\ln\Omega(\beta) & = & 0,\label{samplesbeta}\end{eqnarray}
with \[
n_{k}^{\mathrm{id}}(\beta)=\left\{ \exp\left[-\mu(\beta)\beta+\hbar^{2}k^{2}\beta/2m\right]-1\right\} ^{-1}\]
being the usual Bose-Einstein distribution.

For the purpose of the simulations presented here, a fairly crude
yet effective importance sampling was applied as follows. For relatively
weak coupling $g$, a very rough but useful estimate of the thermal
state at coarse resolution is that the Fourier modes are decoupled
and thermally distributed with some mean occupations $n_{k}(\beta_{t})$
at the target time $\beta_{t}$ that we are interested in. In practice
we will choose some estimate of the guiding density $n_{k}(\beta_{t})$.
The desired equal weight sampling at time $\beta_{t}$ would then
correspond to the distribution \begin{eqnarray}
G^{\text{est}}(\vec{v},\beta_{t}) & = & \delta^{2}(\ln\Omega)\prod_{k}\delta^{2}\left(\widetilde{\alpha}_{k}-(\widetilde{\alpha}_{k}^{+})^{\ast}\right)\notag\\
 &  & \times\frac{\exp[-|\widetilde{\alpha}_{k}|^{2}/n_{k}(\beta_{t})]}{\pi\, n_{k}(\beta_{t})},\label{Gestbeta}\end{eqnarray}
which leads to samples given by $\widetilde{\alpha}_{k}^{(\nu)}=\sqrt{n_{k}(\beta)}\ {\eta_{k}}$
and $\Omega=1$. What we are interested in is the corresponding distribution
of samples at $\beta=0$. An estimate of the initial distribution
that leads to $G^{\text{est}}(\vec{v},\beta_{t})$ can be obtained
by evolving (\ref{Gestbeta}) back in imaginary time using only kinetic
interactions. This is again rather rough, since deterministic interaction
terms $\propto g$ are omitted, not to mention noise, but it is simple
to carry out and proved sufficient for our purposes here. One obtains
then an estimated sampling distribution for samples at $\beta=0$:
\begin{eqnarray}
G^{\text{samp}}(\vec{v},0) & = & \delta^{2}(\ln\Omega-\ln\Omega_{0})\prod_{k}\delta^{2}\left(\widetilde{\alpha}_{k}-(\widetilde{\alpha}_{k}^{+})^{\ast}\right)\notag\\
 &  & \times\frac{\exp(-|\widetilde{\alpha}_{k}|^{2}/n_{k}^{\text{samp}})}{\pi\, n_{k}^{\text{samp}}},\label{Gest0}\end{eqnarray}
where \begin{equation}
n_{k}^{\mathrm{samp}}=n_{k}(\beta_{t})\exp\left[-\lambda-\mu(\beta_{t})\beta_{t}+\frac{\hbar^{2}k^{2}\beta_{t}}{2m}\right],\end{equation}
and the pre-weight $\Omega_{0}\equiv\Omega(0)$ now depends on the
set of particular values of $\widetilde{\alpha}_{k}$ at $\beta=0$
obtained for a given sample, according to \begin{equation}
\ln\Omega_{0}=\sum_{k}|\widetilde{\alpha}_{k}|^{2}\left(\frac{1}{n_{k}^{\mathrm{samp}}}-\frac{1}{\overline{n}_{x}}\right).\end{equation}

For most of the simulations reported here, taking $n_{k}(\beta_{t})$
to be just the ideal gas Bose-Einstein distribution $n_{k}^{\mathrm{id}}(\beta_{t})$
was sufficient. However, once the chemical potential $\mu(\beta_{t})$
approaches or exceeds zero, this estimate is no longer useful. A better
choice for $n_{k}(\beta_{t})$ is the density of states function $\rho_{k}$
of the exact Yang and Yang solution \cite{yangyang1}, although it
should be noted that this is not the density of actual particles that
we seek. In practice, our approach was to first run a calculation
based on this estimate $n_{k}(\beta_{t})=\rho_{k}(\beta_{t})$, obtain
a better estimate of the real density from this full stochastic calculation
by evaluating the expectation value of $\widehat{\Psi}_{k}^{\dagger}\widehat{\Psi}_{k}$
using Eq. (\ref{obs}), then finally use this expectation value to
choose an improved preweighting function $n_{k}(\beta_{t})$ for a
{}``second-generation\textquotedblright\ calculation.

One important point to make regarding the choice of $n_{k}(\beta_{t})$
is that one should endeavor always to choose the preweighting guide
density $n_{k}(\beta_{t})$ equal or greater than the real density,
never smaller. The reasoning behind this is as follows: Suppose first
one chooses a $n_{k}(\beta_{t})$ guiding function that is much smaller
than the true k-space density $n_{k}^{\mathrm{true}}(\beta_{t})$.
This means that the variance of the $\widetilde{\alpha}_{k}$ samples
will be too small to recover the physical value of the density upon
averaging $\langle|\widetilde{\alpha}_{k}|^{2}\Omega\rangle_{\mathcal{S}}$
without resorting to very large weights for the largest $|\widetilde{\alpha}_{k}|$
samples. In practice, if the ratio $n_{k}/n_{k}^{\mathrm{true}}$
is small, then the typical trade-off that occurs is that the largest
contribution to $\Omega|\widetilde{\alpha}_{k}|^{2}$ comes from those
$|\widetilde{\alpha}_{k}|$ that are many standard deviations from
the mean. Their rarity is compensated for by a very large $\Omega$.
However, this is fatal for practical numbers of samples because in
fact not even one of the samples one obtains ends up in this highest-contribution
region at many standard deviations from the mean. For $n_{k}/n_{k}^{\mathrm{true}}\lesssim1/2$,
the number of samples with $|\widetilde{\alpha}_{k}|^{2}\gtrsim n_{k}$
will be $\propto\mathcal{S}\prod_{k}\exp\left[-(n_{k}^{\mathrm{true}}/n_{k})^{2}/2\right)$,
i.e. vanishing, leading to a systematic error.

In contrast, the opposite situation when $n_{k}(\beta_{t})$ is chosen
too large is much more benign. Following the above reasoning, one
gets a distribution of $\widetilde{\alpha}_{k}$ samples that is too
broad, with the result that a majority of samples are too far away
from physical values of $|\widetilde{\alpha}_{k}|^{2}$ and their
excessive abundance must be compensated for by giving them a correspondingly
small weight. However, for reasonably large numbers of trajectories,
there always remains a core of the smallest samples that are in the
region of most important contributions. The number of these samples
is of the order of $\mathcal{S}\prod_{k}n_{k}^{\mathrm{true}}/n_{k}(\beta_{t})$,
which is reasonable in practice as long as the estimate $n_{k}(\beta_{t})$
is not extremely poor.

Finally, it should be mentioned that superior importance sampling
schemes to the crude one we have employed here could be implemented
and may allow one to reach much lower temperatures than presented
here. A first step would be to keep the $\beta=\beta_{t}$ distribution
estimate, Eq. (\ref{Gestbeta}), but estimate the resulting initial
samples at $\beta=0$ in a more accurate manner. To do this, one could
choose the $\beta=\beta_{t}$ samples according to $\widetilde{\alpha}_{k}^{(\nu)}(\beta_{t})=\sqrt{n_{k}(\beta_{t})}\ \eta_{k}$
and $\ln\Omega(\beta_{t})=0$ as usual, but then evolve them back
in time to $\beta=0$ numerically, using the deterministic part of
the full equations (\ref{Gequations}). This would give a superior
estimate of the initial distribution as it takes into account $g\neq0$
mean field effects as well as kinetic evolution. Having these $\beta=0$
samples, one would then proceed forward in time with the full stochastic
evolution.

A further refinement would be to choose initial $\beta=0$ samples
via the Metropolis algorithm, so that the initial samples $\vec{v}$
are distributed according to $\mathcal{F}\left[\vec{v}\right]$, where
$\mathcal{F}=|\Omega(\beta_{t})|$ when $\Omega(\beta_{t})$ is calculated
according to the deterministic part of the evolution, Eq. (\ref{Gequations}),
starting from $\Omega(0)=0$. This avoids the arbitrariness of the
crude Gaussian choice, Eq. (\ref{Gestbeta}). A final, but numerically
intensive approach would be to sample the phase-space variables $\alpha_{j}(\beta_{t})$
and $\text{Im}[{\ln\Omega}](\beta_{t})$ directly via a Monte Carlo
Metropolis algorithm whose free parameters to be varied include both
the initial noises $\eta_{k}$ and all the time-dependent noises $\zeta_{j}^{(\nu)}(\beta)$
for a given time lattice $\beta\in\left(0,\beta_{t}\right)$.

\subsection{Trust indicators for sampling}

\label{append:trust}

One should mention two heuristic trust indicators that we use extensively
to exclude bad sampling of the underlying phase-space distribution.

Firstly, let us point out that the behavior of the evolution equations
(\ref{logequations}) is such that one builds up an approximately
Gaussian distribution of the logarithmic variables (leaving aside
the evolution of $N_{j}$ itself, which is initially small). This
means that the stochastic averages to be evaluated, e.g., in Eq. (\ref{g2obs}),
involve means of \textsl{exponentials} of approximately Gaussian random
variables (as per $\overline{m}=\langle e^{v}\rangle$ with $v$ Gaussian).
A feature of such means is that if the variance of the \textsl{logarithm}
$\text{Re}[v]$ exceeds a value of around $10$ the mean $\overline{m}$
begins to have systematic error when calculated with any practical
sample sizes. This is discussed in detail in \cite{deuar-drummond-pp,deuar-thesis}.
As a result, when calculating observables with some expression $\langle F(\vec{v})\rangle_{\mathcal{S}}$,
one must also check that the variance of its logarithm is small enough,
i.e. that \begin{equation}
\mathcal{V_{F}}=\langle(\ln|F(\vec{v})|)^{2}\rangle_{\mathcal{S}}-\langle\ln|F(\vec{v})|\rangle_{\mathcal{S}}^{2}\lesssim10.\end{equation}
If this is not satisfied, the results for $\langle F(\vec{v})\rangle_{\mathcal{S}}$
must be considered suspect.

Secondly, sampling problems of this sort usually make themselves visible
if one compares two calculations with widely different sample sizes.
In practice one can evaluate an average and its uncertainty with $\mathcal{S}$
samples, and with $\mathcal{S}/10$ samples (where, of course, $\mathcal{S}/10\gg1$).
If the difference is statistically significant the result of the $\mathcal{S}$
sample average again should be considered suspect.

\subsection{Choice of intermediate $\mu(\beta)$}

\label{append:mu}

If one is primarily interested in the behavior of the system around
some target temperature $\beta_{t}$ and chemical potential $\mu(\beta_{t})$
(alternatively -- density), then the values of $\mu(\beta)$ at intermediate
times $\beta<\beta_{t}$ can in principle be chosen at will.

In practice, however, some choices lead to smaller statistical uncertainty
than others because the intermediate values of density affect the
amount of noise generated during the evolution. A preliminary investigation
of $\mu$ choice in \cite{deuar-thesis} indicated some heuristic
guidelines that were also followed in the present work:

(\textit{i}) It is advantageous to not vary $\mu_{e}(\beta)$ too
much over the course of the simulation. Excessive variation leads
to increased noise.

(\textit{ii}) A constant or piecewise-constant value of $\mu_{e}$
is also advantageous because the ideal-gas part of the evolution can
then be calculated exactly in logarithmic variables (\ref{trueequationsk}),
and step-size is only important for the interaction part of the evolution.

(\textit{iii}) It is advantageous to choose an initial density that
is much smaller than the final one at $\beta_{t}$ both for statistical
sampling reasons and because this puts the initial gas much further
into the classical decoherent regime ($\tau\gg\gamma^{2}$), where
the initial condition (\ref{G0}) applies, than the final regime.

In practice, our simulations used the following form \begin{equation}
\mu_{e}(\beta)=\frac{1}{\Delta\beta}\ln\frac{z(\beta+\Delta\beta)}{z(\beta)},\end{equation}
which is piecewise constant over a time step $\Delta\beta$, with
the fugacity \begin{equation}
z(\beta)=e^{\mu\beta}=\left\{ \begin{array}{ll}
z_{i}, & \text{ when }\beta\leq\beta_{i},\\
z_{t}\exp\left[-\frac{\beta_{t}-\beta}{\beta_{t}-\beta_{i}}\ln\frac{z_{t}}{z_{i}}\right], & \text{ when }\beta>\beta_{i}.\end{array}\right..\end{equation}
Here, $\beta_{t}$ and $z_{t}=e^{\mu_{t}\beta_{t}}$ are the target
inverse temperature and fugacity, and $\beta_{i}$ and $z_{i}$ are
numerical constants for the initial high temperature state that we
chose to be $z_{i}^{2}=z_{t}^{2}/1000$ and $\beta_{i}=\beta_{t}/1000$.

Given the difficulty of precisely analyzing the statistical behavior,
it is unclear whether a wiser choice of $\mu(\beta)$ may lead to
significant improvements over the results presented here. However,
this is the most successful choice of those we tried.


\section{Integrals in perturbation theory in $\gamma$}

\label{append:integrals-nearly-ideal}

We begin with Eq. \eqref{eq:weak_first} and substitute the expression
for $\Gamma(k,\sigma)$ in Eq. \eqref{eq:gamma_k_result} to give
\begin{equation}
\Delta g^{(2)}(r)=-\frac{g}{\hbar}\sqrt{\frac{m\beta}{\pi}}\int_{0}^{\beta}d\sigma\frac{\exp\left\{ -\frac{r^{2}m\beta}{4\hbar^{2}[\beta^{2}/4-(\sigma-\beta/2)^{2}]}\right\} }{\sqrt{\beta^{2}/4-(\sigma-\beta/2)^{2}}}.\end{equation}
Next we make the substitution $t=(2/\beta)(\sigma-\beta/2)$ and $y=r\sqrt{m/(\hbar^{2}\beta)}$
to give \begin{eqnarray}
\Delta g^{(2)}(r) & = & -\frac{g}{\hbar}\sqrt{\frac{m\beta}{\pi}}\int_{-1}^{1}dt\frac{e^{-y^{2}/(1-t^{2})}}{\sqrt{1-t^{2}}}\label{appenda_step1.5}\\
 & = & -\frac{g}{\hbar}\sqrt{\frac{m\beta}{\pi}}e^{-y^{2}}\int_{-\infty}^{\infty}dx\frac{e^{-y^{2}x^{2}}}{1+x^{2}},\label{appenda_step2}\end{eqnarray}
where the last equality follows from the substitution $t=x/\sqrt{1+x^{2}}$.
The exponent in the integrand of Eq. \eqref{appenda_step2} can be
represented as a Gaussian integral \begin{equation}
e^{-y^{2}x^{2}}=\frac{1}{\sqrt{\pi}}\int_{-\infty}^{\infty}dke^{-k^{2}+2ikyx}.\end{equation}
Then, changing the order of integration in Eq. \eqref{appenda_step2}
we arrive at \begin{eqnarray}
\Delta g^{(2)}(r) & = & -\frac{g}{\hbar\pi}\sqrt{m\beta}e^{-y^{2}}\int_{-\infty}^{\infty}e^{-k^{2}}\int_{-\infty}^{\infty}\frac{e^{i2kyx}}{1+x^{2}}dxdk\notag\\
 & = & -\frac{2g\sqrt{m\beta}}{\hbar}\int_{0}^{\infty}e^{-(k+|y|)^{2}}dk.\label{appenda_step3}\end{eqnarray}
The final result shown in Eq. \eqref{eq:weak_g2_res} follows trivially
from a shift in the integration variable $k\rightarrow k-|y|$, and
the definition of the complimentary error function, \begin{equation}
\mathrm{erfc}(|y|)\equiv\frac{2}{\sqrt{\pi}}\int_{|y|}^{\infty}dke^{-k^{2}}.\end{equation}


\section{Integrals in the Bogoliubov treatment}

\label{append:Bogoliubov}

We first evaluate the vacuum contribution $G_{0}(r)$, Eq. (\ref{G-0-def}).
Writing down the integral explicitly, in terms of $k$, and transforming
to a new variable $x=k\xi/2$, we have\begin{equation}
G_{0}(r)=\frac{2}{\pi\xi n}\int\limits _{0}^{\infty}dk\left[\frac{x}{\sqrt{1+x^{2}}}-1\right]\cos(2rx/\xi).\end{equation}
Integrating by parts, gives\begin{equation}
G_{0}(r)=-\frac{1}{\pi nr}\int\limits _{0}^{\infty}dx\frac{\sin(2\sqrt{\gamma}nrx)}{(1+x^{2})^{3/2}}.\label{G-0-final}\end{equation}

The integral in (\ref{G-0-final}) can be expressed in terms of special
functions \cite{MathHandbook}, giving \begin{equation}
G_{0}(r)=-\sqrt{\gamma}\left[\mathbf{L}_{-1}(2\sqrt{\gamma}nr)-I_{1}(2\sqrt{\gamma}nr)\right].\end{equation}

The finite temperature term $G_{T}(r)$, Eq. (\ref{G-T-lowT}), is
evaluated by performing variable changes according to $E=$ $\hbar^{2}k^{2}/(2m)$,
followed by $\epsilon=\sqrt{E(E+gn)}$ and then $x=\epsilon/gn$.
In this way we transform the integral over $k$ to an integral over
$x$ \begin{equation}
G_{T}(r)=\sqrt{\frac{2mg}{\pi^{2}\hbar^{2}n}}\int\limits _{0}^{\infty}dx\left[\frac{\sqrt{1+x^{2}}-1}{1+x^{2}}\right]^{1/2}\frac{\cos[k(x)r]}{e^{gnx/T}-1},\label{G_T_a}\end{equation}
where $k(x)=[2mgn(\sqrt{1+x^{2}}-1)/\hbar^{2}]^{1/2}$. So far we
have not made any additional assumptions or approximations.

By inspecting the integrand in Eq. (\ref{G_T_a}) one can see that
for $T\ll gn$ the main contribution to the integral comes from $x\ll1$.
Therefore for $T\ll gn$ ($\tau\ll\gamma$) we can simplify the integral
by treating $x$ in the integrand as a small parameter. Accordingly,
we obtain\begin{gather}
\left[\frac{\sqrt{1+x^{2}}-1}{1+x^{2}}\right]^{1/2}\simeq\frac{1}{\sqrt{2}}x,\; x\ll1,\\
k(x)\simeq\sqrt{\frac{mgn}{\hbar^{2}}}x,\; x\ll1,\end{gather}
and therefore\begin{equation}
G_{T}(r)\simeq\frac{\tau^{2}}{4\pi\gamma^{3/2}}\int\limits _{0}^{\infty}dy\frac{y\cos(\tau nry/2\sqrt{\gamma})}{e^{y}-1},\end{equation}
where we have introduced $y=gnx/T=\epsilon/T$.

Finally we make use of the following integral\begin{equation}
\int_{0}^{\infty}dy\frac{y\cos(ay)}{e^{y}-1}=\frac{1}{2a}-\frac{\pi^{2}}{2}\cosech^{2}\left(\pi a\right),\label{eq:c7}\end{equation}
and obtain Eq. (\ref{G-T-GPa}).

In the opposite limit, dominated by thermal fluctuations and corresponding
to $\gamma\ll\tau\ll1$, we first note that large thermal fluctuations
correspond to $\tilde{n}_{k}\gg1$, which in turn requires $\epsilon_{k}/T\ll1$.
Thus, we replace $\tilde{n}_{k}$ in the integral (\ref{G-T-def})
by $\tilde{n}_{k}=[\exp(\epsilon_{k}/T)-1]^{-1}\simeq T/\epsilon_{k}\gg1$.
As a result, the thermal contribution $G_{T}(r)$ becomes\begin{gather}
G_{T}(r)\simeq\frac{1}{\pi n}\int_{-\infty}^{+\infty}dk\frac{E_{k}T}{\epsilon_{k}^{2}}\cos(kr)\notag\\
=\frac{4mT}{\pi\hbar^{2}n}\int_{0}^{+\infty}dk\frac{\cos(kr)}{k^{2}+(2/\xi)^{2}}=\frac{mT\xi}{\hbar^{2}n}e^{-2r/\xi},\end{gather}
which is valid for $r/\xi\lesssim1$. Rewriting this in terms of the
dimensionless parameters $\gamma$ and $\tau$ we obtain Eq. (\ref{G-T-GPb}).
{For $r/\xi\gg1$ the cosine term becomes important and the values
of momenta in the integral Eq.~(\ref{G_T_a}) are cut off by $1/r\ll\xi$.
In this regime one can use the approximation that led to Eq.~(\ref{eq:c7}).}


\section{Integrals in perturbation theory in $1/\gamma$}

\label{append:integrals-Tonks}

We begin by evaluating the direct contribution given by Eq. \eqref{eq:the_diagram}
by substituting Eq. \eqref{eq:gamma_res}, \begin{eqnarray}
\Delta g_{d}^{(2)} & = & \int_{0}^{\beta}d\sigma\int_{-\infty}^{\infty}\frac{dk}{2\pi}\left(-\frac{2\hbar^{2}k^{2}}{mn\gamma}\right)e^{ikr-\sigma\hbar^{2}k^{2}(\beta-\sigma)/m\beta}\notag\\
 & = & \frac{-1}{\pi\gamma}\sqrt{\frac{\tau}{2}}\int_{0}^{1}ds\int_{-\infty}^{\infty}dq\, q^{2}e^{iqy-sq^{2}(1-s)},\label{appendc_step1.5}\end{eqnarray}
where we have affected the change of variables $\sigma=\beta s$,
$q=\sqrt{\beta\hbar^{2}/m}k$ and $y=\sqrt{m/(\beta\hbar^{2})}r=\sqrt{(\tau n^{2}/2)}r$.
The integration with respect to $q$ can then be done using integration
by parts, which yields \begin{eqnarray}
\Delta g_{d}^{(2)} & = & \frac{-1}{4\gamma}\sqrt{\frac{\tau}{2\pi}}\int_{0}^{1}ds\frac{2s(1-s)-y^{2}}{s^{5/2}(1-s)^{5/2}}e^{-y^{2}/[4s(1-s)]}\notag\\
 & = & \frac{-1}{\gamma}\sqrt{\frac{2\tau}{\pi}}\int_{-1}^{1}dt\left(1-\frac{2y^{2}}{1-t^{2}}\right)\frac{e^{-y^{2}/(1-t^{2})}}{\left(1-t^{2}\right)^{3/2}},\notag\\
 &  & \,\label{appendc_step2}\end{eqnarray}
where the last equality follows from the substitution $s=(t+1)/2$.
The simplest way to solve the integral in Eq. \eqref{appendc_step2}
is by comparison with Eq. \eqref{appenda_step1.5} in Appendix \ref{append:integrals-nearly-ideal}.
In doing so, one may observe \begin{eqnarray}
 &  & \int_{-1}^{1}dt\left(1-\frac{2y^{2}}{1-t^{2}}\right)\frac{\exp\left[-\frac{y^{2}}{1-t^{2}}\right]}{\left(1-t^{2}\right)^{3/2}}\\
 & = & \frac{d^{2}}{dy^{2}}\int_{-1}^{1}dt\frac{\exp\left[-\frac{y^{2}}{1-t^{2}}\right]}{\sqrt{1-t^{2}}}=\pi\frac{d^{2}}{dy^{2}}\text{erfc}(|y|).\end{eqnarray}
The result shown in Eq. \eqref{eq:direct_exchange} then follows trivially
from this.

In order to calculate the exchange contribution we begin with Eq.
\eqref{delta_g_exchange} and substitute Eq. \eqref{eq:gamma_res},
which immediately yields \begin{equation}
\Delta g_{e}^{(2)}(r)=\frac{1}{\gamma}\sqrt{\frac{\pi\tau}{2}}e^{-in\tau r^{2}/2}F_{e}(\sqrt{\tau n^{2}r^{2}/2})\end{equation}
where $F_{e}(y)=\int_{0}^{1}ds\int dq\, q^{2}e^{-s(1-s)q^{2}+i(1-2s)qy}/\pi^{3/2}$,
and $s$, $q$ and $y$ are defined the same was as for the direct
contribution. The integration with respect to $q$ can be carried
out using integration by parts, leaving an integral with respect to
$s$: \begin{eqnarray}
 &  & \int_{0}^{1}ds\frac{\exp\left[-\frac{y^{2}(1-2s)^{2}}{4s(1-s)}\right]}{s^{3/2}(1-s)^{3/2}}\left[1-\frac{y^{2}(1-2s)^{2}}{2s(1-s)}\right]\notag\\
 & = & 4\int_{-1}^{1}dv\frac{\exp\left[-\frac{y^{2}v^{2}}{1-v^{2}}\right]}{(1-v^{2})^{3/2}}\left[1-\frac{2v^{2}y^{2}}{1-v^{2}}\right]\notag\\
 & = & 4\int_{-\infty}^{\infty}dt\left[1-2y^{2}t^{2}\right]e^{-y^{2}t^{2}}\end{eqnarray}
where the first equality comes from the substitution $s=(v+1)/2$
and the second from $v=t/\sqrt{1+t^{2}}$. Both terms are standard
definite integrals it is straightforward to show that \begin{equation}
\Delta g_{e}^{(2)}=\frac{4}{n\gamma}\delta(r).\end{equation}
Thus the only effect of the exchange contribution is to cancel the
delta-function contribution coming from the direct contribution at
$r=0$.



\begin{thebibliography}{84}
\bibitem{note-Wick} Note that Eq. \eqref{eq:wick} only contains 4 terms coming from Wick's
theorem, the other 20 terms are disconnected corrections (in the
language of Feynman diagrams) and hence only produce corrections to
the single particle Green functions. That is they represent the
interacting corrections to the relation between chemical potential
and density.
\end{thebibliography}
\end{document}